\documentclass[emulateapj,epsfig]{article} 
\usepackage[onecolumn]{emulateapj}
\usepackage[]{epsfig}
\hoffset -1.5truecm \lefthead{Menci et al.} \righthead{ } 

\begin{document}

\title{BINARY AGGREGATIONS IN HIERARCHICAL GALAXY FORMATION: \\ THE EVOLUTION OF 
THE GALAXY LUMINOSITY FUNCTION} 
\author{N. Menci}
\affil{Osservatorio Astronomico di Roma,
 via Osservatorio, 00040 Monteporzio, Italy}
\author{A. Cavaliere}
\affil{Astrofisica, Dipartimento Fisica, II Universit\`a di Roma,\\ via Ricerca 
Scientifica 1, 00133 Roma, Italy} 
\author{A. Fontana, E. Giallongo, F. Poli} 
\affil{Osservatorio Astronomico di Roma,  via Osservatorio, 00040 Monteporzio, 
Italy}
 
\smallskip
\begin{abstract}
We develop a semi-analytic model of hierarchical galaxy formation with 
an improved treatment of the evolution of galaxies inside dark matter 
haloes. We take into account not only dynamical friction processes 
building up the central dominant galaxy, but also binary aggregations 
of satellite galaxies inside a common halo. These deplete small to 
intermediate mass objects, affecting the slope of the luminosity 
function at its faint end, with significant observable consequences. 
We model the effect of two-body aggregations using the kinetic 
Smoluchowski equation. This flattens the mass function by an amount 
which depends on the histories of the host haloes as they grow by 
hierarchical clustering. The description of gas cooling, star 
formation and evolution, and Supernova feedback follows the standard 
prescriptions widely used in semi-analytic modelling. 
We find that binary aggregations are effective in depleting the number of 
small/intermediate mass galaxies over the redshift range $1<z<3$, 
thus flattening the slope of the luminosity function at the faint 
end. At $z\approx 0$ the flattening occurs for $-20<M_B<-18$, but an 
upturn is obtained at the very faint end for $M_B>-16$. We compare our 
predicted  luminosity functions with those obtained from deep 
multicolor surveys in the HDF-N, HDF-S, NTT-DF in the rest-frame B and 
UV bands for the redshift ranges $0<z<1$ and $2.5<z<3.5$, 
respectively. The comparison shows that the discrepancy of the 
predictions of other semi-analytic models with the observations  
is considerably reduced at $z>1$ and even more at 
$z\approx 3$ by the effect of binary aggregations. 
The predictions from our dynamical model are discussed and 
compared with the effects of complementary processes (additional 
starburst recipes, alternative sources of feedback, different mass 
distribution of the dark matter haloes) which may conspire in 
affecting the shape of the luminosity function.
\end{abstract}

\section{INTRODUCTION}

Understanding of galaxy formation has undergone impressive developments in 
recent years. On the observational side, a key step in the study of the 
statistical properties of high-redshift galaxies has been taken with the 
discovery of the "Lyman-break'' galaxies at $z\approx 3-4$ (Steidel et al. 1996; 
Adelberger et al. 1998); other important steps have been taken with the estimate 
of the cosmic star formation rate (see Madau, Pozzetti \& Dickinson 1998; 
Fontana et al. 1999), and the measurement of the galaxy luminosity functions 
(LFs) at redshifts up to $z\approx 4$ for UV luminosities extending over the 
range $-24\lesssim M\lesssim -18$ (Steidel et al. 1999 from spectroscopic data, 
Pozzetti et al. 1998 and Poli et al. 2001 from photometric redshifts). 

These results, with their unprecedented span in both cosmic time and magnitude, 
boosted the developement of an ``ab initio'' theory of galaxy formation. The 
theory has been firmly rooted into the cosmological framework with the 
development of semi-analytic models (SAMs, Kauffmann et al. 1993; Cole et al. 
1994; Somerville \& Primack 1999; Poli et al. 1999; Wu, Fabian \& Nulsen 2000; 
Cole et al. 2000) which link in a single computational structure two main 
blocks: the dynamical history of galaxies as they emerge from primordial dark 
matter (DM) density perturbations and grow through hierarchical merging events; 
the baryonic processes in the galactic structures, namely, gas cooling, star 
formation, rise and fall of stellar populations, energy feedback from 
Supernovae.

Such advances in both theory and observations are pushing the comparison between 
models and data to higher and higher degrees of accuracy. A key role is played 
by the $z$-resolved LFs. In fact, the earliest SAMs were calibrated to fit the 
locally observed LF (still with considerable uncertainties in the normalization 
and in the slope at the faint end), and were tested at higher redshift through 
integrated counts and $z$-distributions. The recent observations are producing a 
progressive convergence of the local LFs found by different  groups (see Zucca 
et al. 1997, Cross et al. 2001) down to $M_B\approx -16$, with the additional 
indication that at the faint end the shape is more complex than represented by a 
Schechter form (see Marzke, Huchra \& Geller 1994; Loveday 1997). In addition, 
the resolved LFs now begin to describe the evolution of the galaxy population 
out to $z\approx 4$, thus providing a {\it differential} test for the model 
predictions.

The comparison of the models with such data supports the grand design of 
hierarchical galaxy formation, but indicates that some of the processes included 
in the SAMs require an improved treatement. For example, the  LFs predicted at 
$z\approx 3$ overestimate the number of faint ($M_{1700}\gtrsim -17$) galaxies 
by a factor $5-8$ when compared with observations based on photometric reshifts 
(see Somerville, Primack \& Faber 2001; Poli et al. 2001), and the excess goes 
beyond the estimated incompleteness of the data faintwards of $m_B\sim 26$. A 
similar trend, though less evident, is found in the redshift range around 
$z\approx 1$ (Poli et al. 2001), while at $z\approx 0$ the observed upturn in 
the LF at the faint is not accounted for in the current SAM predictions. 
Addressing the above critical points is crucial to fully understand the physical 
processes driving the galaxy  evolution. 

The shape of the predicted LF at faint and intermediate luminosities is affected 
by two main processes.  The first concerns  the effect of Supernovae (SN 
hereafter) which heat and partially expell the galactic gas. A stronger SN 
feedback would suppress star formation in smaller haloes thus decreasing their B 
and UV luminosity and flattening the LF at the faint end. However, in the 
framework of the simple parametrizations currently adopted in SAMs it is 
extremely difficult to increase such feedback without destroying the agreement 
of the model with other observables; in particular, the predicted luminosities 
for small spiral galaxies would be too faint when compared with existing data 
concerning the Tully-Fisher relation (see Cole et al. 2000). Attempts at non-
parametric treatements of the feedback have been made by Mac Low \& Ferrara 
(1999); Goodwin, Pearce \& Thomas (2000); Monaco (2001). Different sources of 
feedback, like that arising from the photoionization of the inter-galactic 
medium by stars and quasars, have been recently included in the SAM framework
(Benson et al. 2001).

The other component affecting the shape of the LF is the mass distribution of 
the galaxies, which is  determined by the detailed dynamical processes taking 
place inside the host DM haloes. Among these, tidal stripping and binary 
aggregations of satellite galaxies play a relevant role. However, the former 
affects mainly the very low-mass end of the mass distribution at circular 
velocities $v\approx 20-50$ km/s, and has a minor impact on the faint end of the LF 
(see Benson et al. 2001);  so we shall focus here on binary aggregations. Such a 
process is treated in current SAMs only under the approximation of orbital decay 
toward a central dominant galaxy. Here we extend the treatement to include 
aggregations between all galaxies in common DM haloes. On the other hand, we 
shall adopt the standard SAM prescriptions concerning star formation to derive 
the galaxy LF from the mass distribution; no additional starburst recipes are 
associated with the aggregation events between satellite galaxies. This allows 
us to single out the dynamical effects of our description without introducing 
new (and uncertain) free parameters in the model, as we discuss in \S 6. 

The paper is organized as follows: the derivation of the mass function in the 
framework of SAMs, and the motivation for our improved treatment are discussed 
in \S 2. A technical description of our approach is presented in \S 3, where we 
show how we fit our treatment of binary aggregations into the canonical SAM 
framework. In \S 4 we briefly recall the basic prescriptions that we share with 
other SAMs to  correlate the gas cooling, the star formation and evolution, and 
the SN feedback with the dynamical history of the galaxies. The LFs we derive 
are compared with the data in \S 5. In \S 6 we discuss our results and present 
our conclusions.

\section{THE DYNAMICAL EVOLUTION OF GALAXIES: OVERVIEW AND SPECIFIC 
MOTIVATIONS}

In SAMs the galaxy mass distribution is derived from the merging histories of 
the host DM haloes, under the assumption that the galaxies contained in each 
halo coalesce into a central dominant galaxy if their dynamical friction 
timescale is shorter than the halo survival time; the surviving galaxies 
(commonly referred to as satellite galaxies) retain their identity and continue 
to orbit within the halo. While the histories of the DM condensations rely on a 
well established framework (the extended Press \& Schechter theory, EPST, see 
Bower 1991; Bond et al. 1991; Lacey \& Cole 1993), the recipe concerning the 
galaxy fate inside the DM haloes is guided by {\it a posteriori} consistency 
with the outputs of high-resolution  N-body simulations.

In reality, additional dynamical processes complement the dynamical friction in 
driving the evolution of the mass distribution. Among these, binary aggregations 
between satellite galaxies in common haloes have previously been considered by 
Cavaliere, Colafrancesco \& Menci (1991, 1992); Cavaliere, \& Menci (1993), as a 
process that  would flatten the shape of the mass function (and hence of the LF) 
at small/intermediate masses. This is because such masses aggregate into larger 
units, while the large masses are so few that their binary encounters are 
unlikely. In the above papers, the aggregation-driven evolution of the mass 
distribution was computed in terms of the {\it non-linear} Smoluchowski kinetic 
equation with a mass-dependent aggregation rate. The results showed that the 
effectiveness of the process depends critically on the environment, which in 
those models was a given input. To describe at the same time the galaxy dynamics 
and the evolution of the host haloes by hierarchical clustering, high-resolution 
N-body simulations or SAMs are required.

On the N-body side, recent works (see Klypin et al. 1999 for pure DM, and Murali 
et al. 2001 for hydrodynamical simulations) indicate a complex galaxy growth. 
While at very low circular velocities ($v\approx 20-50$ km/s) tidal stripping may 
affect the mass distribution (Gnedin \& Ostriker 1997; see also Taylor \& Babul 
2000; Taffoni et al. 2001), at larger masses ($v\sim 100$ km/s) binary 
aggregations of satellite galaxies do play a relevant role which complements 
the coalescence into a central galaxy through dynamical friction. Indeed, Murali 
et al. (2001) analyze the competing role of the two growth modes of the 
simulated galaxies in terms of an evolutionary equation which includes also a 
binary aggregation term of the Smoluchowski type.

On the SAM side, efforts to insert the aggregations between satellite 
galaxies have recently been started by Somerville \& Primack (1999). 
However, they adopt a cross section (derived from the 
N-body simulations of galaxy encounters by Makino \& Hut 1997) valid for 
equal galaxies in clusters with velocity dispersion much higher than the 
internal galaxy  dispersion, a condition that allowed them to adopt a one-
body treatement for the aggregations.

Here we take a step forward, and consider in closer detail the actual {\it 
two-body} dynamics of aggregations; we shall also adopt a cross section 
valid down to small groups with velocity dispersions close to those 
internal to galaxies, as is the case at early times in the hierarchical 
clustering picture.

In fact, we develop a SAM including both dynamical friction 
and binary aggregations. Instead of employing a Monte Carlo simulation, 
as usual for SAMs, we follow the evolution of the galaxy mass distribution 
by solving numerically a set of evolutionary equations, as in Poli et 
al. (1999). The subset describing the two-body dynamics is constituted by the 
(non-linear) kinetic Smoluchowski equation. This modifies the mass function 
resulting from dynamical friction by an amount which depends on the properties 
of the host DM haloes, which in turn evolve according to the EPST theory. The 
basic description of gas cooling, star formation and evolution, and SN feedback 
is kept unchanged with respect to the standard SAM prescriptions given, e.g., by 
Poli et al. (1999) and Cole et al. (2000).

We first compare our results with those from existing N-body simulations, and 
then compare our LFs with those obtained from spectroscopic and deep imaging 
surveys down to faint magnitutes ($I_{AB}\approx 27.2$), at redshifts up to 
$z\approx 4$. This allows us to test the effects of binary aggregations on the 
SAM predictions over a wide range of cosmic times and masses. 

\section{THE GROWTH OF THE GALAXY MASSES}

We consider the number $N(m,M,t)\,dm\,dM$ per Mpc$^3$ of galaxies with mass in 
the range $dm$ about $m$, residing in haloes with masses in the range $M$ to 
$M+dM$ at cosmic time $t$. At the initial time $t_0$ we assign one galaxy to 
each halo, a condition which formally translates into 
$N(m,M,t_0)=N_H(M,t_0)\delta(m-M)$ where $\delta$ is the Dirac delta function. 
Our default choice for the halo mass distribution $N_H(M,t)$ is the standard 
Press \& Schechter (1974) expression, whose dependence on the cosmological 
parameters and on the spectrum of primordial density perturbation is  recalled 
in Appendix A; but we shall also explore the effects of adopting the Sheth \& 
Tormen (1999) mass distribution, also recalled in Appendix A.

Given the merging history of DM haloes described by EPST, two main processes 
affect the evolution of $N(m,M,t)$: the orbital decay of satellite galaxies onto 
a central dominant galaxy due to dynamical friction, and the binary aggregations 
between satellite galaxies. In the following two subsections we investigate the 
effects of the above processes on the evolution of $N(m,M,t)$.  A schematic view 
of the two dynamical processes driving the growth of galaxies in our model is 
given in fig. 1. The corresponding evolution of the mass distribution of 
galaxies is presented in the last subsection 3.4. 
\vspace{-0.2cm}
\begin{center} 
\scalebox{0.46}[0.43]{\rotatebox{0}{\includegraphics{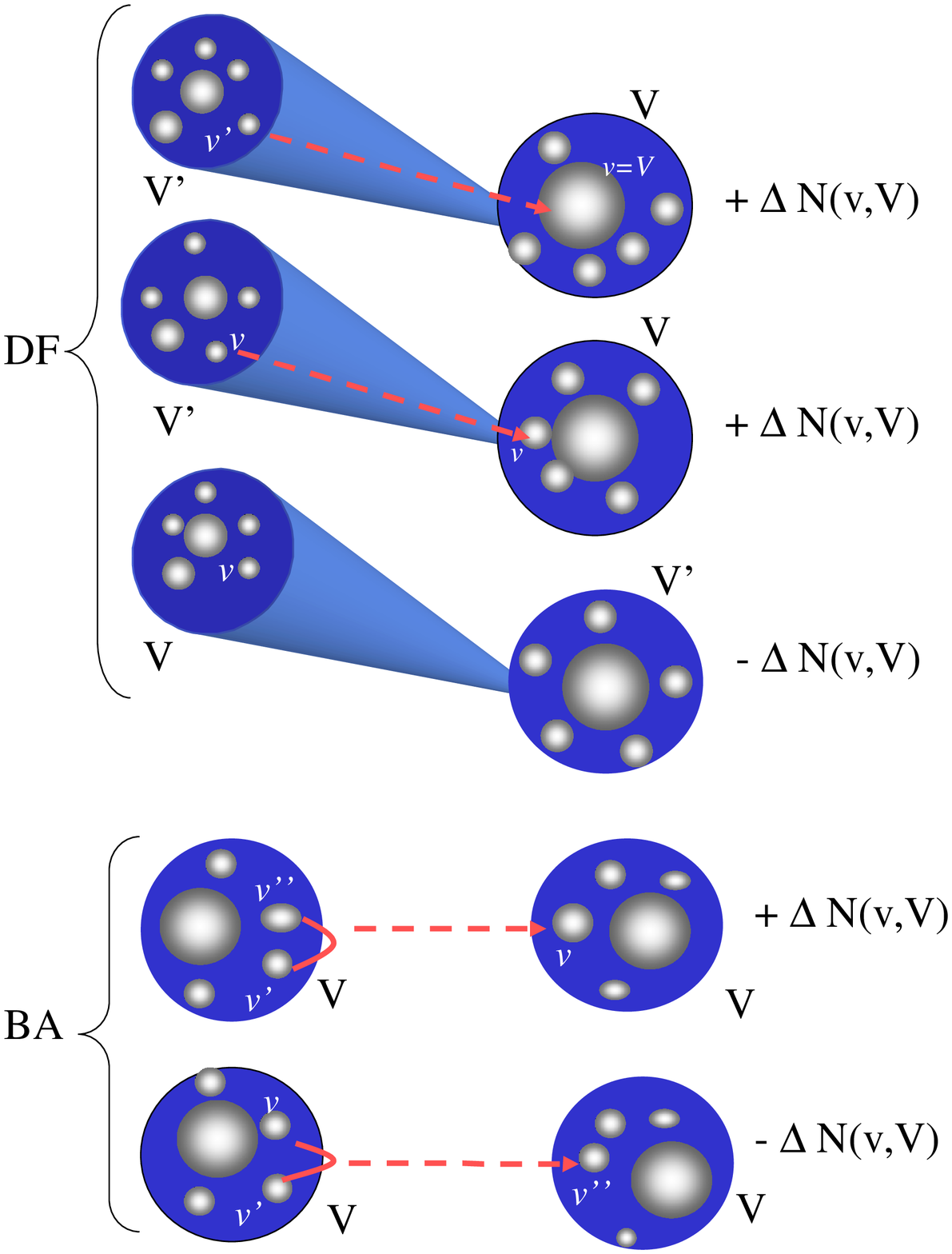}}}
\end{center}
{\footnotesize
\vspace{-0.3cm } 
Fig. 1. - A schematic representation of the various terms contributing to the 
evolution of the galaxy velocity function $N(v,V)$ inside DM haloes of circular 
velocity $V$. The first three processes marked with DF correspond to the 
construction and destruction terms (eq. 1) from dynamical friction following the 
merging of the host haloes (see text). The last two processes, marked with BA, 
represent binary aggregations of satellite galaxies (see eq. 3) inside the DM 
host halo. Their combined effects drive the evolution of $N(v,V,t)$. 
\vspace{0.2cm}}

\subsection{Coalescence driven by dynamical friction}

In the following we shall frequently adopt the circular velocities as 
the proper quantity to mark the depth of the potential wells of 
galaxies and of their host haloes. As for the latter, this is related to 
the mass through the relation $V=\sqrt{GM/R}$ where the limiting 
radius $R$ is the radius within which the mean mass density is 
$200\,\rho_c$, where $\rho_c$ is the critical density at the redhsift 
$z$ where the halo is identified. The relation between $R$ and $M$ is 
given, e.g., in Navarro, Frenk \& White (1997) and Mo, Mao \& White 
(1998) and takes the form 
$R=1.63\,10^{-2}\,(M/h^{-1}\,M_{\odot})^{1/3}\,
[\Omega_0/\Omega(z)]^{-1/3}\,(1+z)^{-1}\,h^{-1}\,kpc$ 
for Einstein-de-Sitter ($\Omega=1$, $\Omega_{\Lambda}=0$), 
open ($\Omega_0<1$, $\Omega_{\Lambda}=0$), and flat 
($\Omega_0+\Omega_{\Lambda}=1$) universes. 

As to a galaxy inside a host DM halo, the circular velocity $v$ is 
related to $m$ through $m=v^2\,r_{tid}/G$;  
here $r_{tid}$ is the tidal radius, within which the mean density of 
the galaxy exceeds the average density of the host halo interior the 
pericentre of the galactic orbit. This ensures that the galactic 
subhalo survives the tidal stripping due to the host halo potential wells 
and retains its own identity. Our computation of $r_{tid}$ is 
given in Appendix B. 

The above relations allow us to relate the mass and the circular 
velocity distributions by applying the proper Jacobian.

The evolution of $N(v,V,t)$ driven by dynamical friction is computed as follows. 
At each time step, we first compute the  conditional probability $d^2 
P_H(V'\rightarrow V,t)/d V'\,dt$ that a given halo with circular velocity $V$ at 
time $t$ has a progenitor with circular velocity $V'$ at $t'=t-\Delta t$; this 
is calculated in the framework of the EPST, and its expression (depending on 
cosmology and on the perturbation spectrum as given in Bower 1991;  Bond et al. 
1991; White \& Frenk 1991, Lacey \& Cole 1993), is recalled in Appendix A for a 
generic $t'$ (see eq. A-3). Similarly, from the EPST we compute the inverse 
conditional probability $d^2 P_H(V\rightarrow V',t)/d V'\,dt$ that a given halo 
with circular velocity $V$ at time $t$ ends up in a halo with circular velocity 
$V'$ at $t+\Delta t$ (see eq. A-2).

According to the canonical prescriptions usually adopted by SAMs, during halo 
merging a galaxy contained in one of the parent halos contributes to enrich the 
dominant galaxy of the common halo of circular velocity $V$ if its coalescence 
time $\tau_{df} (v)$ is shorter than the halo survival time $\tau_l(V)$ 
predicted in the EPST. The circular velocity of the merger is then equated to 
that of the host halo. Or else, for $\tau_{df}>\tau_l$, the galaxy retains its 
identity. Thus the increment in the number $N(v,V,t)$ of galaxies of given $v$ 
in halos of given $V$ is linked to the increment of DM halos with $V$, whose 
contribution from progenitors $V'$ is $N_H(V',t)\, d^2\,P_H(V'\rightarrow V,t)/ 
d V' d t$; the link occurs through the probabilities: i) of forming one dominant 
galaxy of circular velocity $v=V$ by coalescing  galaxies contained in the 
parent halos with lower velocities $v'<v$; ii) of finding galaxies with 
velocities $v$ which have not coalesced despite the merging of their host 
haloes. The corresponding decrement is due to the inclusion of galaxies with 
current velocity $v$ into a larger halo $V'>V$. The construction and destruction 
terms described above are schematically represented in the upper part of fig. 1. 
The corresponding evolution of $N(v,V,t)$ in a single timestep $\Delta t$ is 
expressed by 
\begin{eqnarray}\label{dynfrict} 
& & N(v,V,t+\Delta t) - N(v,V,t) = 
\nonumber  \\ & = & \Delta t\,\delta(v-V)\,\int_0^{v}\,dv'\, \int_{v'}^{V}\,dV' 
\,N_H(V',t)\, { d P_H(V'\rightarrow V,t)\over  d V' d t}\, {N(v',V')\over 
N_{T}(V')} \,prob\big[\tau_{df}(v')<\tau_l (V)\big] \nonumber \\ &+& \Delta 
t\,\int_{v}^{V}\,dV' \,N_H(V',t)\,{ d^2 P_H(V'\rightarrow V,t)\over  d V' d t}\, 
{N(v,V')\over N_{T}(V')}\,\Big\{1-prob\big[\tau_{df}(v)<\tau_l (V)\big]\Big\} 
\nonumber\\ &-& \Delta t\,\int_{V}^{\infty}\,dV' { d^2 P_H(V\rightarrow 
V',t)\over  d V' d t}\,N(v,V) ~,
\end{eqnarray} 
where $N_{T}(V)=\int dv'\,N(v',V,t)$ 
is the total number of galaxies per $\rm Mpc^3$, per unit halo 
rotational velocity. The first term on the right-hand side implies that if any 
galaxy with circular velocity $v'<V$ is included in a halo with circular 
velocity $V$ and its orbit decays (due to dynamical friction) to the center of 
such a halo within the halo survival time, then a central galaxy with velocity 
$v=V$ is formed inside the halo.

Coalescence is caused by loss of galaxy energy and orbital angular momentum due 
to dynamical friction to the halo material. The timescale of such process is 
usually determined through orbit averaging of the Chandrasekhar formula 
(see Lacey \& Cole 1993); the result is given in Cole et al. (2000) as follows: 
\begin{equation}\label{dftime} 
\tau_{df}=\Theta\,\,\tau_{dyn}\,{0.3722\over ln(M/m)}\,{M\over m}~.
\end{equation}
Here $\tau_{dyn}\equiv \pi\,R/V$ is the dynamical time of the halo, and 
$\Theta=[J/J_c(E)]^{0.78}\,[r_c(R)/R]^2$ contains the dependence on the initial 
energy $E$ and angular momentum $J$ of the galaxies, in terms of the angular 
momentum $J_c$ and the radius $r_c$ of the circular orbits corresponding to $E$. 
Although the values of $\Theta$ are statistically distributed (approximately 
following a lognormal function with $\langle log_{10}\,\Theta\rangle=-0.14$ and 
$\langle (log_{10}\,\Theta-\langle log_{10}\,\Theta\rangle)^2 \rangle^{1/2} 
\approx 0.26$, see Tormen 1997), in the following we shall use for $\Theta$ its 
average value. Note that the orbit averaging procedure (adopted to derive eq. 2) 
is not appropriate for very elongated orbits for which the local density varies 
on a much shorter timescale than that given by eq. (2); however, such orbits are 
expected to be a minority ($\lesssim 15\%
$, see Ghigna et al. 1998).

The probability $prob\big[\tau <\tau_l (V)\big]$ for a DM halo of velocity $V$ 
to have a survival time $\tau_l$ larger than a given value $\tau$ has been 
computed by Lacey \& Cole (1993, see their eq. 2.21) in the framework of the 
EPST, and is recalled in Appendix A.

\newpage
\subsection{Binary aggregations of satellite galaxies}

In addition to the above coalescence process, we include binary aggregations. So 
for each DM halo with circular velocity $V$, we compute the further evolution of 
the galaxy mass distribution in a timestep due to binary aggregations. This is 
described by the Smoluchowski equation, which we write in terms of masses for 
the sake of simplicity: 
\begin{eqnarray}\label{smoluch} 
N(m,M,t+\Delta t) & - & N(m,M,t) =  \nonumber \\ 
& & {1\over 2}\,\Delta t\,\int_0^{m}\,dm'\,N(m',M,t)\,N(m-m',M,t)\,\tau^{-
1}_{agg}(m',m-m',V) \nonumber \\ &-& \Delta
t\,N(m,M,t)\,\int_0^{\infty}\,dm'\,N(m',M,t)\,\tau^{-1}_{agg}(m,m',V)
~, 
\end{eqnarray}
where $\tau^{-1}_{agg}(m,m',V)$ is the aggregation rate depending on the DM halo 
where the galaxies $m$ and $m'$ reside. The first term describes the 
construction of galaxies with mass $m$ from smaller ones with mass $m'$ and $m-
m'$, while the second represents the destruction of galaxies $m$ due to their 
aggregation with others. A schematic representation of the two terms governing 
the binary aggregations is given in the lower part of fig. 1.

The aggregation rate is governed by $\tau^{-1}_{agg}=\Sigma\,V_{rel}/(4\pi 
R^3/3)$ (see Cavaliere, Colafrancesco \& Menci 1992). Here $V_{rel}$ is the 
average relative velocity of galaxies in the DM halo whose rms value is equal to 
twice the halo 1-D velocity dispersion $\sigma_V\approx V/\sqrt{2}$, and 
$\Sigma$ is the cross section. The latter is given for nearly grazing, weakly 
hyperbolic encounters by Saslaw (1985), and by Cavaliere, Colafrancesco, \& 
Menci (1992). It includes a geometrical term (proportional to the area of the 
galaxies), and a focussing factor $\sim 1+(v/V_{rel})^2$ that accounts for the 
enhancement of $\Sigma$ in slow encounters with resonance between the internal 
and the orbital degrees of freedom  (see also Binney \& Tremaine 1986). Thus, 
the average rate for binary aggregations is 
\begin{equation}\label{aggtime}
\tau^{-1}_{agg}=\Big\langle
\,\pi\,(r^2+r'^2)\, \Big(1+G{m+m'\over r+r'}{1\over V^2_{rel}}\Big)\,
{V_{rel}\over 4\pi R^3/3} \Big\rangle~, 
\end{equation}
where the average is over the relative velocities $V_{rel}$. The distribution of 
the encounter velocities is assumed to be Maxwellian, namely 
\begin{equation}\label{maxwell}
g(V_{rel})=\sqrt{2\over \pi}\,{\,V_{rel}^2\over(\sqrt{2}\,\sigma_V)^3}\,
e^{-V_{rel}^2/4\sigma_V^2}~.
\end{equation}

Note that for encounters between equal galaxies with $r'=r$ and $N(M')=N(M)=N$ 
the scaling of the aggregation rate (\ref{aggtime}) reduces to $\tau_{agg}^{-
1}\propto \langle r^2\,(1+v^2/V_{rel}^2)\,V_{rel}\rangle$. Performing the 
average over the distribution $g(V_{rel})$ in terms of the rescaled variable 
$y\equiv V_{rel}/v$ yields $\tau_{agg}^{-1}\propto r^2\,v^4\,\sigma_V^{-3}\,R^{-
3}\,I(x)$, where the function $I(x)\equiv \int\,dy\,y^3\,exp(-y^2/x^2)+ 
\int\,dy\,y\,exp(-y^2/x^2)$ tends to a constant when the ratio $x\equiv 
\sigma_V/v\rightarrow \infty$. Thus the aggregation rate (\ref{aggtime}) reduces 
to the expression given by Makino \& Hut (1997) -- originally derived analytically 
by Mamon (1992) -- and the r.h.s. of eq. (3) 
becomes proportional to $N^2\,r^2\,v^4\,\sigma_V^{-3}\,R^{-3}$ (the effective 
rate adopted by Somerville \& Primack 1999) in the proper limits of large 
encounter velocities relative to the internal galaxy velocity dispersion, and of 
encounters between equal galaxies. 
 
Finally, note that after performing the average of eq. (4) over 
the distribution $g(V_{rel})$ only $\sigma_V$ enters the computation;   
in other words, we use an average description for encounter velocities 
typical of the environment considered. The geometrical
cross section contained in eq. (4) does not 
properly describe single events with $V_{rel}\gg v$, but these 
are rare for our typical values of $\sigma_V/v$ 
in the range from $1$ to about $4$. In cases with larger ratios 
$\sigma_V/v$ the effect of the aggregations is suppressed 
since $\tau_{agg}\propto \sigma_V^3$ as shown above, 
but even inside rich clusters the averaged cross section 
that we adopt is consistent with the simulation 
results, see Makino \& Hut (1997); for numerical computations of the 
role of interactions inside clusters see also Lanzoni (2000).  

In sum, our cross section provides an accurate 
{\it average} description of aggregations in systems where the 
velocity dispersion 
is close to the galaxy circular velocity and the focussing term is relevant; 
on the other hand, it constitutes an effective average approximation for 
encounters in environments 
with large velocity dispersion up to the scale of rich clusters, where 
the aggregations are disfavoured anyway. 
As a final global check, we have verified that the insertion of an artificial cutoff 
at $V_{rel}=4\,v$ in $\Sigma$ does not change our results. 

\subsection{Numerical solutions of the equations: test cases}

The equations (\ref{dynfrict}) and (\ref{smoluch}) describing the 
evolution of $N(m,M,t)$ are integrated numerically on a grid of 
circular velocities and cosmic times with step 
$\Delta t=10^{-2}\,H_o^{-1}$. 

In order to test our numerical code, we first run the computation for 
two relevant simple cases where analytic solutions are available. In 
the limit $\tau_{df}\rightarrow 0$ (i.e., when 
merging of the host haloes is promptly followed by 
coalescence of the galaxies within them by dynamical friction) 
the solution $N(v,V,t)$ of 
eq. (1) when  integrated over the circular velocity $V$ must yield the 
Press \& Schechter mass distribution. The comparison between the 
numerical and the analytic solution in this case is performed at three 
different times in the top panel of fig. 2, which shows that the 
numerical solutions remain close to the analytic Press \& Schechter 
form over the whole range of $v$; in fact, the relative deviation is 
always smaller than 5\%. 

To test the code section concerning the Smoluchowski equation, we numerically 
solve eq. (3) in the case of constant aggregation rate 
$\tau_{agg}^{-1}={\rm const}$ for galaxies within a 
host halo of given mass $M$. 
In this case the exact solution 
(Smoluchowski 1916; Trubnikov 1971) has the form 
$N(m,t)=[A_o/m_*^2(t)]\,exp[-m/m_*(t)]$, where the constant $A_o$ is 
related to the total mass ${\mathcal  M}$ contained in the system, and 
$m_*(t)=m_{*0}\,(t/t_0)$ is a characteristic mass linearly growing 
with time. Trubnikov (1971) has shown 
that such a solution holds for very general initial conditions after a 
transient time. The comparison with the numerical solution is 
performed in the bottom panel of fig. 2, where the initial 
condition (dotted line) has been chosen to be $N(m,t_0)\propto 
(m/m_{*0})^{-1}\,exp[-m/m_{*0}]$ normalized as to yield a
total mass ${\cal M}=1.54\,10^2\,m_{*0}$. Note how the numerical solution 
progressively flattens at small masses during the transient, to 
approach the exact solution (computed at $t=3\,t_0$) 
with a relative error smaller than 
$3\%$ over the whole mass range. Note also that the 
numerical solution conserves the total galaxy mass  
with high accuracy. 

\begin{center} \scalebox{0.65}[0.48]{\rotatebox{0}{\includegraphics
{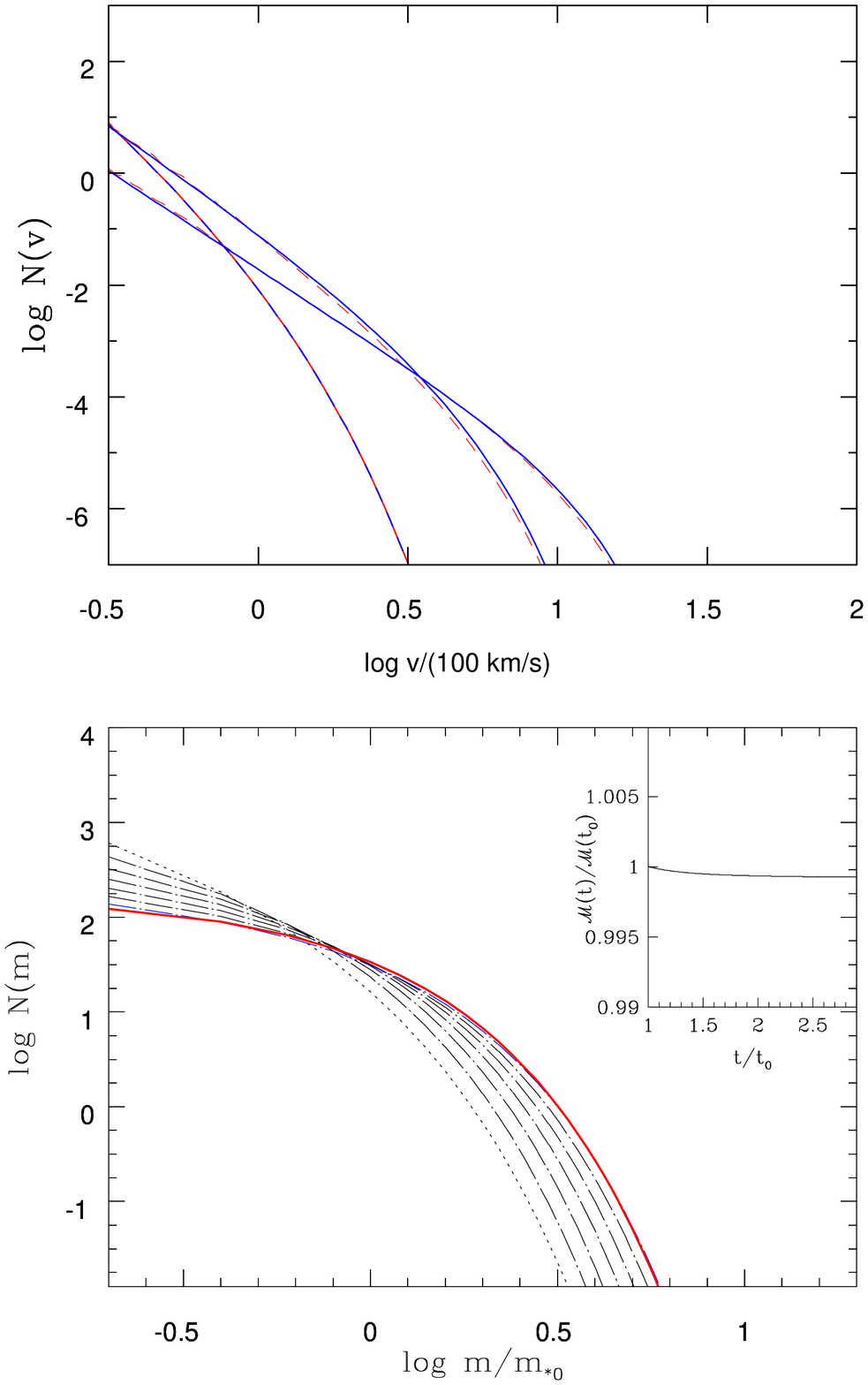}}}
\end{center}
{\footnotesize
Fig. 2 - Comparison between the numerical and the analytic solutions of eqs. (1) 
and (3) for two test cases. Top panel: the numerical solutions of eq. (1) in the 
limit of vanishing $\tau_{df}$ (dashed curves) are compared with the Press \& 
Schechter form (solid curves). The curves refer to redshifts $z=8$, 2.6 and 0 
(from left to right at the high-mass end).  Bottom panel: the numerical solution 
of the Smoluchowski eq. (3), computed for a constant aggregation time (dot-
dashed curves) at 6 equally spaced time intervals from $t_0$ to $3\,t_0$, are 
compared with the corresponding analytic solution (given in the text) computed 
at $t=3\,t_0$ (heavy solid line). The dotted line represents the initial 
condition. The inset shows the variation with time of the total 
mass $\mathcal{M}$ corresponding to the numerical solution. 
} 
\vspace{0.4cm}

\subsection{The evolution of the galaxy circular velocity distribution}

Having tested our code, we proceed to compute the complete evolution of 
$N(v,V,t)$. At each time step, we first compute the change of $N(v,V,t)$ due to 
dynamical friction (the right hand side of eq. \ref{dynfrict}), and then the 
further change due to binary aggregations of satellites (the right hand side of 
eq. \ref{smoluch}) using the physical values of $\tau_{df}$ and $\tau_{agg}$ 
given in eqs. (2) and (4). A discussion of the role of the two terms is given in 
the final \S 6.

Once the complete evolution of $N(v,V,t)$ is found for all times, we compute the 
probability of finding a galaxy with circular velocity $v$ in a halo of circular 
velocity $V$ as $f(v,V,t)\,dv =N(v,V,t)\,dv/N_H(V,t)$. Then the total density 
distribution of galaxies with circular velocity $v$ is computed from 
$N(v,t)=\int_{v}^{\infty}\,dV\,N_H(V,t)\,f(v,V,t)$, where the halo distribution 
$N_H(V,t)$ takes the canonical Press \& Schechter form. This is the distribution 
of circular velocities irrespective of the halo to which the galaxies belong 
(the global velocity distribution).

The ``transition probability'' for galaxies is given by 
$p(v',t',v,t)=\int_{v'}^{\infty}\int_{v}^{\infty} \,dV\,dV'\,{d 
P_H(V',t'\rightarrow,V,t)\over d V'} \,f(v',V',t')\,f(v,V,t)$, where $d 
P_H(V',t'\rightarrow,V,t) \over d V'$ is the fraction of mass in haloes of 
circular velocity $V'$ at time $t'$, and later at $t>t'$ in haloes with $V$; 
this is given by EPST (see Appendix A).  

The evolution of the galaxy circular velocity distribution $N(v,t)$ resulting 
from the full dynamics is shown in fig. 3 for a CDM power spectrum of primordial 
density perturbations and for our reference set of cosmological/cosmogonical 
parameters: $\Omega_0=0.3$, $\Omega_{\lambda}=0.7$, and Hubble constant $h=0.7$ 
in units of 100 km/s/Mpc. For comparison, we also show the evolution 
corresponding to dynamical friction alone (as usually considered in SAMs), and 
the evolution of the halo velocity distribution as given by the Press \& 
Schechter formula.

\begin{center} \scalebox{0.75}[0.7]{\rotatebox{0}{\includegraphics
{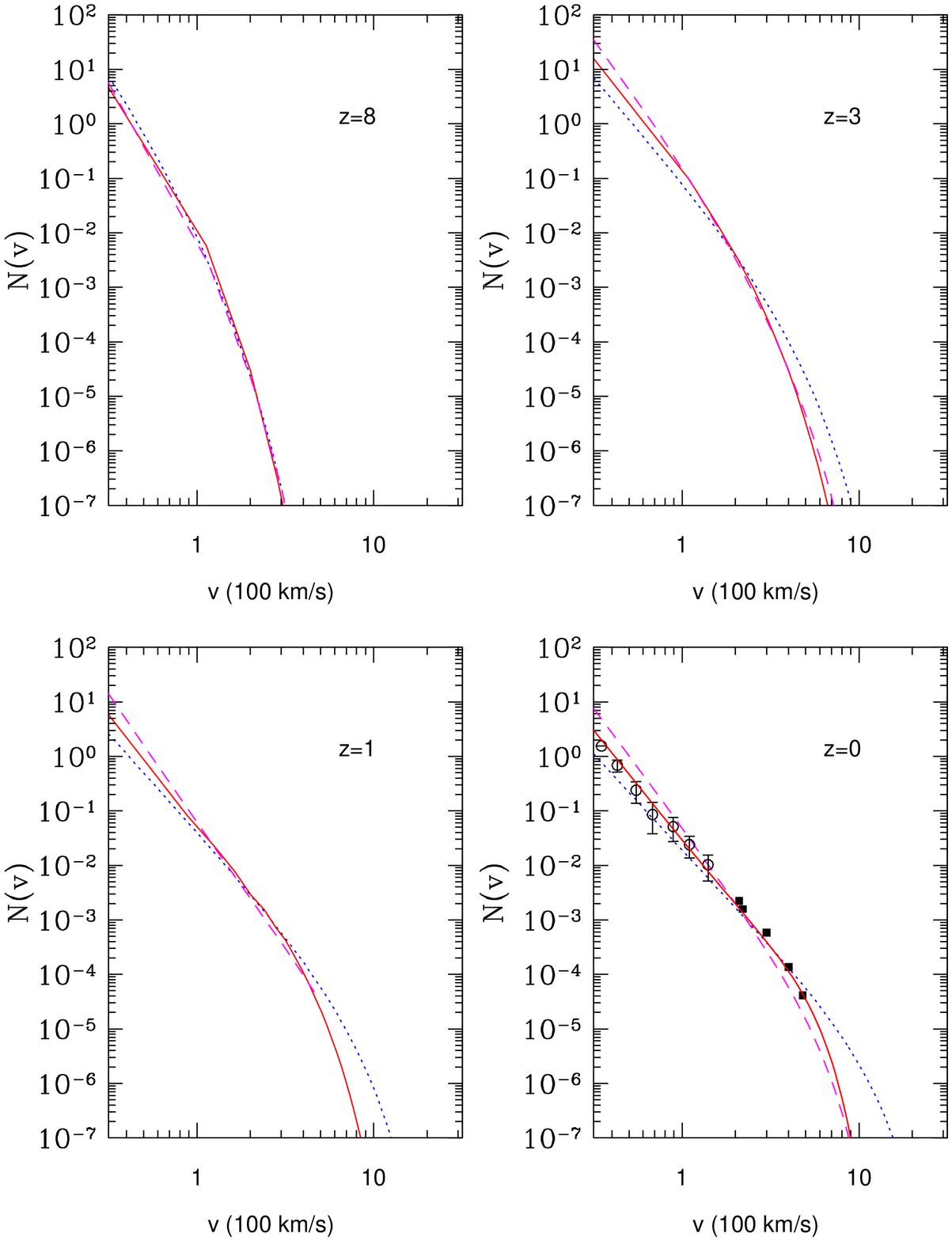}}}
\end{center}
{\footnotesize
Fig. 3 -  
The galaxy velocity function at four different redshifts. The velocity 
is measured in units of 100 km/s. The solid line is the velocity function 
resulting from the full model including dynamical friction and satellite 
aggregation (eqs. 1 and 3). The function $N(v)$ resulting from dynamical 
friction alone is shown as a dashed line, while the velocity function of the 
host haloes (computed from the Press \& Schechter formula) is also shown for 
comparison as a dotted line. In the bottom-right panel we also show the velocity 
function resulting from the N-body simulations by Klypin et al. (1999); the 
solid squares correspond to a simulation with a box size of $60\,h^{-1}$ Mpc, 
and the open circles to a box size of $7.5\,h^{-1}$ Mpc which allows for a finer 
mass resolution. Both cases are computed for $\Lambda$-CDM cosmology with parameters 
as in the text.
\vspace{0.3cm}}

First consider the effect of dynamical friction. Compared to the halo 
velocity distributions, the galaxy distribution shows a delayed 
evolution for $z<3$, particulary evident at low redshift. This is due 
to the longer timescale on which dynamical friction occurs (eq. 
\ref{dftime}) as compared to the survival time of the halo (which is 
of order $\tau_{dyn}$). As a consequence, while at high redshift the 
merging of haloes is promptly followed by coalescence of the galaxies 
within them, at later times the number of galaxies accumulating in the 
haloes increases due to the longer time taken by dynamical friction to 
occur. At small masses this implies more galaxies than haloes, while 
at large $v$ the increase in the number of haloes is not followed by a 
corresponding increase of massive galaxies since these are formed on a 
longer timescale.

The effect of binary aggregations is to flatten the slope of the galaxy velocity 
distribution at the low-mass end ($v\sim 50-150$ km/s). Such a process is 
already effective at $z\approx 2.5$ and continues down to $z\approx 0$, where 
our $N(v)$ is fully consistent with available results from N-body simulations as 
shown by fig. 3. However, in the range from $z\approx 1$ to $z\approx 0$, the 
aggregations deplete more efficiently objects with intermediate mass ($v\sim 
100-150$ km/s), thus producing an upturn of the local velocity distributions at 
low $v$. This is because at low $z$ the galaxies are hosted in haloes that are 
typically more massive than at higher $z$, according to the hierarchical 
clustering. The large relative velocities $V_{rel}$ that galaxies have in such 
deeper potential wells suppress the aggregation of galaxies with low $v$ due to 
the $v/V_{rel}$ term in the cross sections (\ref{aggtime}), as stressed by 
Cavaliere \& Menci (1997). 

To further assess  the consistency of our results with N-body simulations and to 
guide our interpretation of the aggregation role we plot in fig. 4 the overall 
merging rates predicted by our model and compare them with those obtained in 
recent hydrodynamical simulations in a cosmological framework (Murali et al. 
2001). In the top panel we show the net number increment as a function of the 
cosmic time for all galaxies with baryonic mass larger than 
$5.6\,10^{10}\,M_{\odot}$ (the baryonic mass associated with the DM mass in our 
model will be derived in \S 4). We also show as a dashed line the construction 
rate (i.e., the sum of the positive terms in eq. [\ref{dynfrict}] and 
[\ref{smoluch}], also illustrated in fig. 1), to show that it dominates over the 
destruction terms for such galactic  masses. 
\begin{center} 
\scalebox{0.6}[0.48]{\rotatebox{0}{\includegraphics
{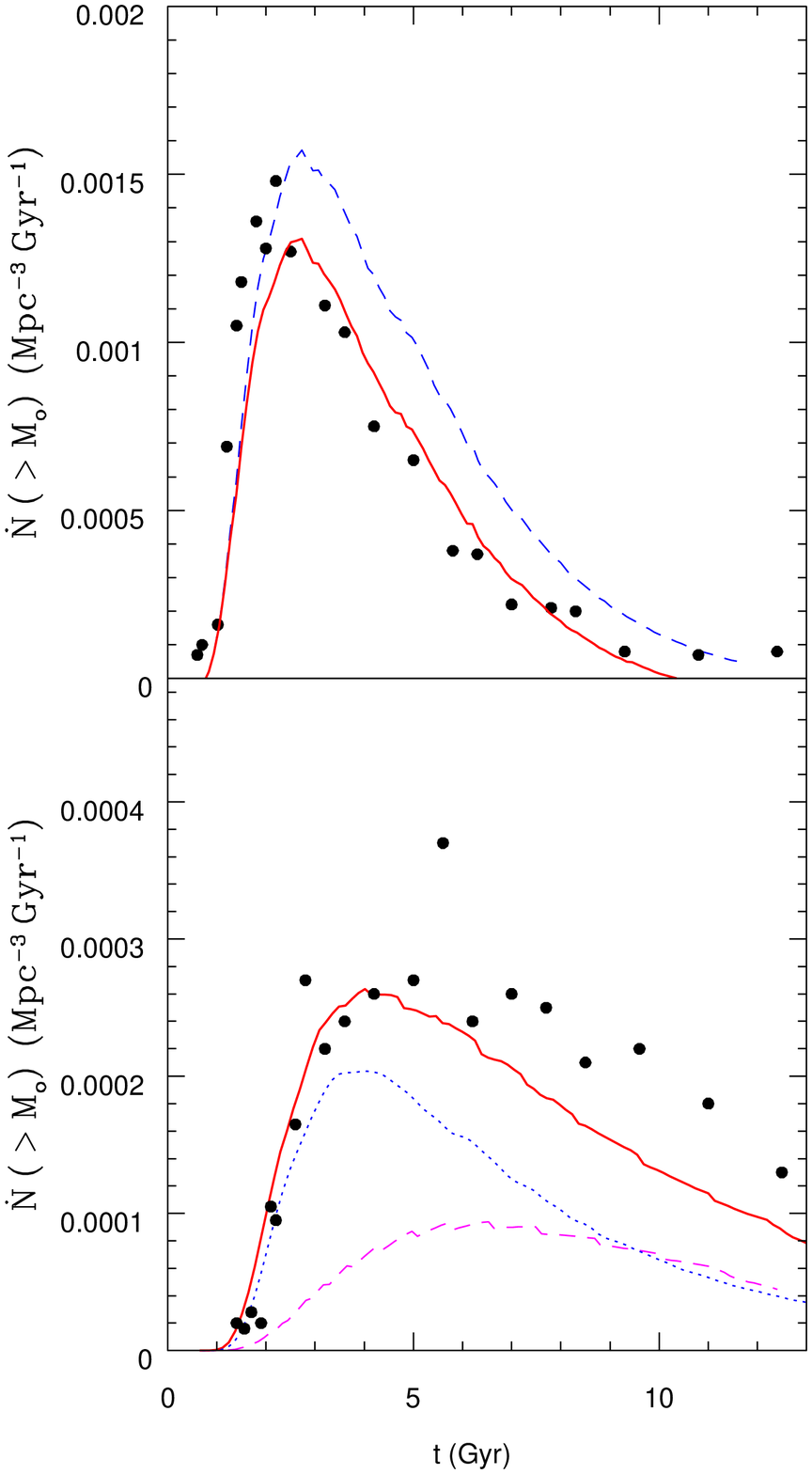}}}
\end{center}
{\footnotesize
Fig. 4 - 
Contributions to the number density of galaxies with masses above 
$5.6\,10^{10}\,M_{\odot}$. Top panel: the net rate (creation minus destruction) 
predicted by the full model including dynamical friction and aggregations of 
satellite galaxies is shown as a solid line. This is compared to that obtained 
from the simulations by Murali et al. (2001, represented by the dots) with the 
same cosmological parameters adopted here. The dashed line shows the 
construction rate alone, as predicted by the model. Bottom panel: the 
destruction rate (the difference between the two curves in the top panel) 
resulting from the model (solid heavy line) is compared with the simulations 
(dots). The contributions to such destruction rate from dynamical friction 
(dotted line) and from aggregations of satellite (dashed line) are also shown.
\vspace{0.2cm}}

Notice how simulations and our model agree in yielding a net rate which declines 
with time after a peak at $z\approx 2.5-3$. The decline of such a rate for 
$z\lesssim 1$ is consistent with the observed one (see LeFevre et al. 2000; 
Carlberg et al. 2000) within the uncertainties entering the comparison.

To assess the relative role of dynamical friction and satellite aggregations we 
show in the bottom panel the destruction rate for the same mass threshold, 
compared with the N-body results. Note the smooth decline at low redshift in 
both the simulations and the model (the heavy solid line) after the peak at 
$z\approx 2.5$. Fig. 4 shows in detail that the position of the peak is mainly 
determined by dynamical friction; satellite aggregations 
begin to contribute significantly to the destruction rate at $z\approx 2.5-3$. 
Then they sustain the destruction rate during its slow decline to lower $z$. 
Thus, first the hierarchical clustering drives the construction (top panel) and 
the destruction (bottom panel) of the basic galactic blocks within a short 
stretch of cosmic time. Then binary aggregations further deplete the number of 
galaxies. The evolution of the destruction rate corresponding to binary 
aggregations is due to the shifting balance of different terms: first, the 
aggregation efficiency increases due to the growth of the galaxy sizes and of 
the number of galaxies contained in the host halos (see the aggregation rate in 
eq. \ref{aggtime}); at later times, the above terms are balanced and/or 
overwhelmed by the increase of the galaxy relative velocities, which grow in 
time following the increasing mass of the host haloes and tend to suppress the 
binary aggregation rate eq. (\ref{aggtime}).

Such a picture is confirmed by the fact that when a lower mass threshold is 
considered, the high-$z$ shape of the destruction rate remains similar to that 
in fig. 4, while the low-$z$ tail is appreciably suppressed. This is because for 
small galaxies the effect of the increase of the relative velocities with time 
is stronger, and prevents binary aggregations from sustaining the 
destruction rate at small $z$. So at $z\approx 0$ the velocity functions in fig. 
3 has an upturn at small masses.

Having discussed the full dynamics, we now turn to recall the basic points of 
the stellar section of our SAM.

\smallskip
\section{STAR FORMATION AND EVOLUTION}

To link the stellar section of the model with the dynamics we adopt the standard 
procedure commonly used in SAMs (Somerville \& Primack 1999; Cole et al. 2000); 
we shall give here a brief presentation of how such a procedure fits into our 
statistical treatement of the evolution of the mass distribution of galaxies 
inside host haloes.   

The baryonic content $(\Omega_b/\Omega_m)\,m$ of the galaxy is divided into a 
hot phase with mass $m_h$ at the virial temperature $T=(1/2)\,\mu\,m_H\,v^2/k$ 
($m_H$ is the proton mass and $\mu$ is the mean molecular weight), a cold phase 
with mass $m_c$ able to radiatively cool within the galaxy survival time, and 
the stars (with total mass $m_*$) forming from the cold phase on a time scale 
$\tau_*$. 

Initially, all the baryons (for which we adopt a density parameter 
$\Omega_b=0.02$) are assigned to the hot phase. Then, for each galaxy mass $m$ 
(with circular velocity $v$ and radius $r$) we compute the baryons in the three 
phases as follows. 

a) {\it The cold phase}. Its mass $m_c$ is increased from inside out by cooling 
processes: at each time step $\Delta t$, we have then $\Delta m_c(v)= 
4\pi\,r_{cool}^2\,\rho_g\,\Delta r_{cool}$; for the gas density we adopt the 
form $\rho_g\propto 1/(r^2+r_{core}^2)$, with the value of $r_{core}$ determined 
by requiring that the density at the virial radius is the same that would have 
been obtained has no gas cooled (hence depending on the past history of $m_c$ 
and $m_h$ corresponding to each galaxy mass $m$ at time $t$). The cooling radius 
$r_{cool}$ at time $t$ is computed by equating the cooling time 
$\tau_{cool}=(3/2)\,kT/\mu\,m_H\,\rho_g(r)\,\Lambda (T)$ to the current time 
$t$. The cooling function $\Lambda (T)$ is taken from Sutherland \& Dopita 
(1993) for a mixture of 77 \% 
hydrogen and 23 \% helium.

The gas which cools settles into a rotationally supported disk; following 
Mo, Mao \& White (1998), its radius $r_d(v)$ and the rotation velocity $v_d(v)$ 
are related to the galaxy circular 
velocity assuming that the angular momentum of the baryons which settle into the 
disc is a fixed fractions $j_d$ of the total angular momentum $J$ of the galaxy 
dark halo.  
The latter is usually expressed in the adimensional form 
$\lambda\equiv J/(|E|^{1/2}/G\,m^{5/2})$. We use the relations 
$r_d(v,\lambda,c,m_d,j_d)$ and $v_d(v,\lambda,c,m_d,j_d)$ given by the 
above authors for a fraction of cold gas $m_d=m_c/m$ in a 
Navarro, Frenk \& White (1997) potential with concentration $c(v)$. In the 
following, such relations will be used only to compute the Tully-Fisher relation 
resulting from our model (see fig. 5 below) and the dust extinction to be 
applied to the galaxy integrated stellar emission (see eq. \ref{sed} and below). 
In such computations we will assume $j_d=0.05$ (as discussed by the above 
authors) and integrate over the full distribution of $\lambda$ (a 
lognorm expression with mean $\overline{\lambda}=0.05$, and dispersion 
$\sigma_{\lambda}=0.5$ in $log(\lambda)$, see Warren et al. 1992; Cole \& Lacey 
1996; Steinmetz \& Bartelmann 1995) to obtain average values for $v_d$ and $r_d$. 
The disk sizes resulting from the above values of $j_d$ and $m_d$ are consistent 
with recent observations, as shown in Giallongo et al. (2000). 

b) {\it The stars}: the mass of baryons turned into stars 
in the timestep $\Delta t$ is assumed to be  
$\Delta m_*=\Delta t\,m_c/\tau_*$, with a star formation timescale 
$\tau_*=\epsilon_*^{-1}\,\tau_{d}\,(v_d/200\,{\rm km/s})^{\alpha_*}$ 
proportional to the dynamical 
time of the disk $\tau_{d}\equiv r_d/v_{d}$. The normalization $\epsilon_*$ and 
the exponent $\alpha_*$ are free parameters. Our default choice for them is 
$\epsilon_*=0.05$ and $\alpha_*=1.5$, as in the fiducial model of Cole et al. 
(2000). This allows us to directly compare ours with previous results, and to 
single out the effects of the dynamics on the LFs. Note that the star formation 
is linked only to the changes of the cold material following the 
dynamical history of the galaxies; as discussed in \S 6 we do not include 
possible starbursts associated with newly formed mergers. 

c) {\it The hot phase} is replenished by part of the cool baryons, namely, 
those re-heated and ejected into the hot phase by SN feedback. This is the most 
uncertain among the {\it baryonic} processes included in SAMs, and is currently 
parametrized by the simple expression $\Delta m_h=\Delta 
m_*\,(v/v_h)^{\alpha_h}$ relating the amount   $\Delta m_h$ of the mass re-
heated in a time step $\Delta t$ to the mass of stars $\Delta m_*$ formed. The 
exponent $\alpha_h$, a free parameter, models the increased feedback efficiency 
in galaxies with decreasing mass.  The choice of $\alpha_h$ has a considerable 
impact on the faint end of the LF (larger values yielding flatter shapes), but 
its values are strongly constrained by the observed Tully-Fisher relation. In 
fact, increasing $\alpha_h$ produces fainter luminosities for small mass 
objects.  This flattens the resulting LF at the faint end, but it also moves the 
predicted Tully-Fisher relation at small $v$ away from the observed range. Our 
default choice of the parameters is $\alpha_h=2$ and $v_h=200$ km ${\rm s}^{-1}$, 
since these yield the best joint fit to both the local LF and the Tully-
Fisher relation, as shown below.

After having computed $\Delta m_c$, $\Delta m_*$ and $\Delta m_h$ from the 
processes a), b), c), we compute the increments due to galaxy 
coalescence/aggregation. The probability $p(v',t',v,t)$ that a galaxy 
with circular velocity $v'$ at time $t'$ is included into a galaxy with velocity 
$v$ at time $t$ has been calculated in \S 3.4. Then, the average 
star content in a halo with circular velocity $v$ is updated over the time grid 
according to following equation: 
\begin{eqnarray} 
m_*(v,t)=\sum_{i}
\int_0^{v}\,dv' \,{N(v',t_i)\over N(v,t)} \,p(v',t_i,v,t)\,\Delta 
m_*(v',t_i)~, 
\end{eqnarray} 
where the sum extends over the time steps $t_i=t_0+i\cdot\Delta t$ ranging from 
the initial time $t_0$ to $t$. Analogous equations govern the increments of $m_c$ 
and $m_h$. Such a procedure is similar to that introduced by Frenk \& White 
(1991); however, here the transition probability for the galaxies is computed on 
the basis of the dynamics described in \S 2.

From the latter equation, the corresponding integrated stellar emission 
$S_{\lambda}(v,t)$ at the wavelength $\lambda$ is computed by convolving with 
the spectral energy distribution $\phi_{\lambda}$ obtained from population 
synthesis models: 
\begin{equation}\label{sed}
S_{\lambda}(v,t) = \int_0^t\,dt'\,\phi_{\lambda}(t-t')\,\dot m_*(v,t')~. 
\end{equation}
Note that the average star formation $\dot m_*(v,t')$ of galaxies with circular 
velocity $v$ at $t'$ is that corresponding to the star mass computed in eq. (6). 
Substituting its expression into eq. (7) 
demonstrates that $S_{\lambda}(v,t)$  contains the 
integrated contributions of star formation in the smaller progenitor systems 
(with circular velocity $v'$ at times $t'<t$) which by the time $t$ 
have been included in the haloes with circular velocity $v$.  The integrations 
over the time $t'$ and the velocity $v'$ entering eq. (7) account 
for the average building up of the stellar population in hierarchically growing 
galaxies, as described by Frenk \& White (1991).

In the following we adopt $\phi_{\lambda}$ taken from Bruzual \& Charlot (1993), 
with a Salpeter IMF. The dust extinction affecting the above luminosities is 
computed assuming the dust optical depth to be proportional to the metallicity 
$Z_{cold}$ of the cold phase (computed assuming a constant effective yield and 
that the metals are re-ejected to the hot phase in the same proportion as the 
reheated gas $\Delta m_h$) and to the disk surface density, so that for the $V$-
band $\tau_{V}\propto m_c\,Z_{cold}/\pi\,r_d^2$. The proportionality constant is 
taken as a free parameter chosen as to fit the bright end of the local LF (see 
below and fig. 5); this yields for the proportionality constant the value 
$3.5\,M_{\odot}^{-1}\,{\rm pc}^2$ when the stellar yield is such as to produce a 
solar metallicity for a $v=220$ km/s galaxy. To compute the extinction in the 
other bands different extinction curves will be considered, including the Milky 
Way (MW), the Small Magellanic Cloud (SMC), and the Calzetti extinctions, see 
Calzetti (1997).
\vspace{-0.5cm}
\begin{center} 
\scalebox{0.35}[0.35]{\rotatebox{0}{\includegraphics{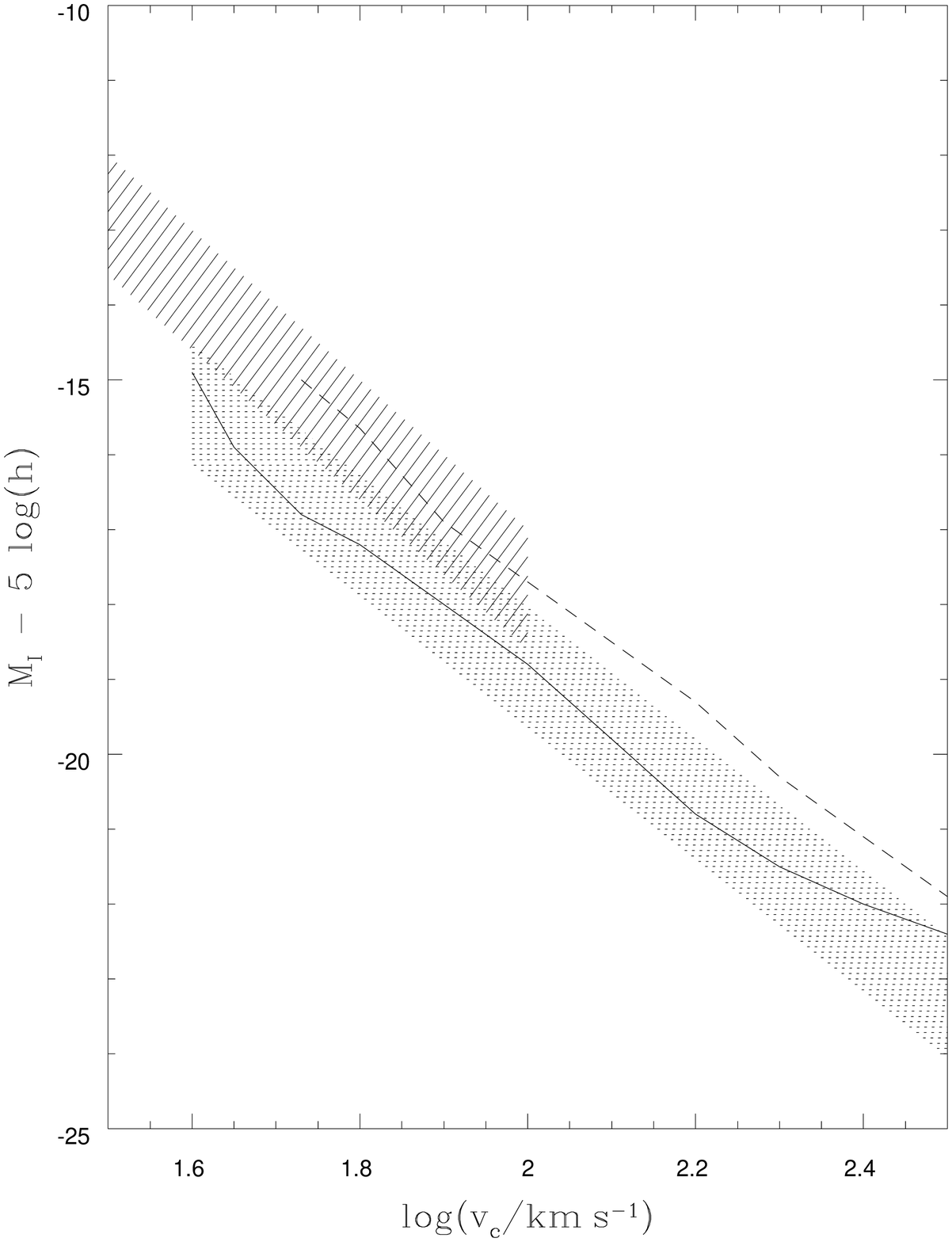}}} 
\scalebox{0.35}[0.35]{\rotatebox{0}{\includegraphics{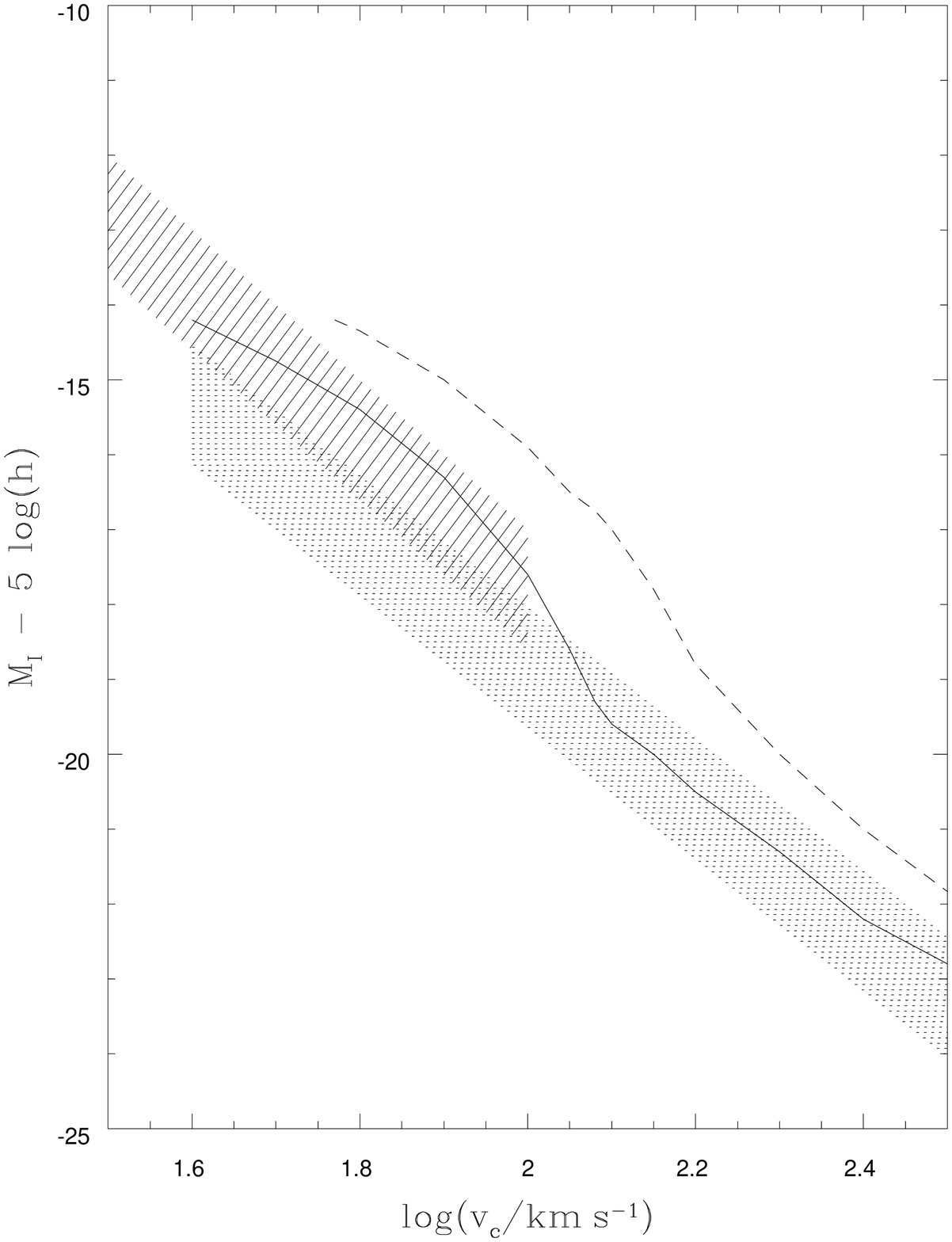}}}
\end{center}
\footnotesize{
Fig. 5 -  
The Tully-Fisher relation relating the disk rotation velocity to the 
luminosity of spiral galaxies. Left panel refers to the our fiducial case 
$\alpha_h=2$, while the right panel to the high-feedback case $\alpha_h=5$. The 
shaded areas represent the region of the $M_I-v$ plane allowed by the 
observations (Mathewson et al. 1992; Willick et al. 1996; Giovanelli et al. 
1997). The solid curves represent the model predictions assuming the disk 
rotation velocity equal to the DM circular velocity $v$, while the dashed curves 
are computed adopting the disk circular velocity $v_d(v)$ as a measure of the 
disk rotation velocity. 
\vspace{0.2cm}} \normalsize

The above equation (7) relates the observable properties of galaxies 
(specifically their integrated stellar emission $S_{\lambda}$) to their 
dynamical properties. The luminosity-circular velocity relation constitutes a 
first, key prediction that can be tested against the observed Tully-Fisher 
relation. The comparison between the outcomes of our model (with our default 
choices for the free parameters) and the observations is shown in the top panel 
of fig. 5. The agreement is satisfactory, although the predicted magnitudes at 
given $v$ are slighly fainter than the data when the disk circular velocity 
$v_d(v)$ is used as a measure of the disk rotation velocity, as appropriate.

To show that increasing $\alpha_h$ is not a viable way to get flatter LFs (as 
anticipated above at point c), we also show the Tully-Fisher relation with 
$\alpha_h=5$, the value initially adopted by Cole et al. (1994) to get a flat 
local LF; in this case the predictions at faint luminosities fail to match the 
observations. Other

\section{THE EVOLUTION OF THE GALAXY LUMINOSITY FUNCTION}

From the galaxy velocity distribution $N(v,t)$ derived in \S 3 (see fig. 3) and 
the $S_{\lambda}-v$ relation discussed above we derive the galaxy LF predicted 
by our model. The results are compared with the observational LF obtained from 
deep surveys. In particular, our predictions at high $z$ are compared with the 
observational LFs derived by Poli et al. (2001) using photometric redshifts for 
galaxies in the ESO New Technology Telescope deep field and in the Hubble Deep 
Field North and South, down to the limiting  magnitudes $I_{AB}=27.2$.  The LFs 
have been computed in the rest-frame B band for $0<z<1$, and at the  rest-frame 
wavelength of 1700 {\AA} for higher $z$. We stress that in the magnitude range 
where the photometric data overlap with those from spectroscopic surveys (see, 
e.g., Steidel et al. 1999), the photometric and spectroscopic LFs agree to a 
remarkable degree of precision.

\subsection{Comparison with data: the effect of satellite aggregations}

The comparison between the observed and the predicted LF is given in figs. 6 and 
7. We compare with the data both the LFs correponding to the standard model 
(where only coalescence driven by dynamical friction is present) and those from 
the complete dynamics including binary aggregations between satellite galaxies; 
hereafter we shall refer to such cases as DF and DF+BA. In both cases, we adopt 
our reference set of cosmological parameters (given in \S 3.4) for a Universe 
dominated by a cosmological constant.

\begin{center} 
\scalebox{0.42}[0.42]{\rotatebox{0}{\includegraphics{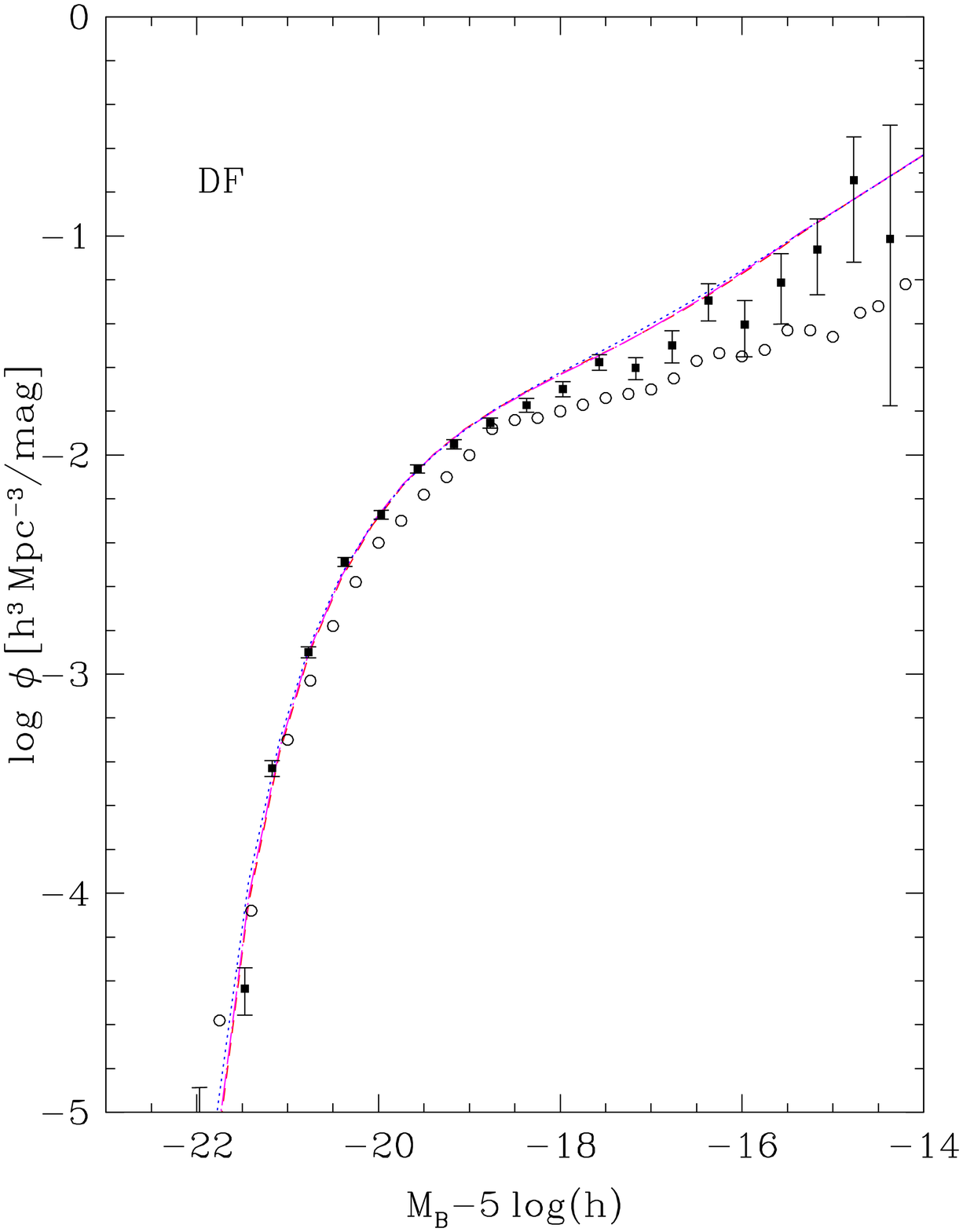}}} 
\scalebox{0.42}[0.42]{\rotatebox{0}{\includegraphics{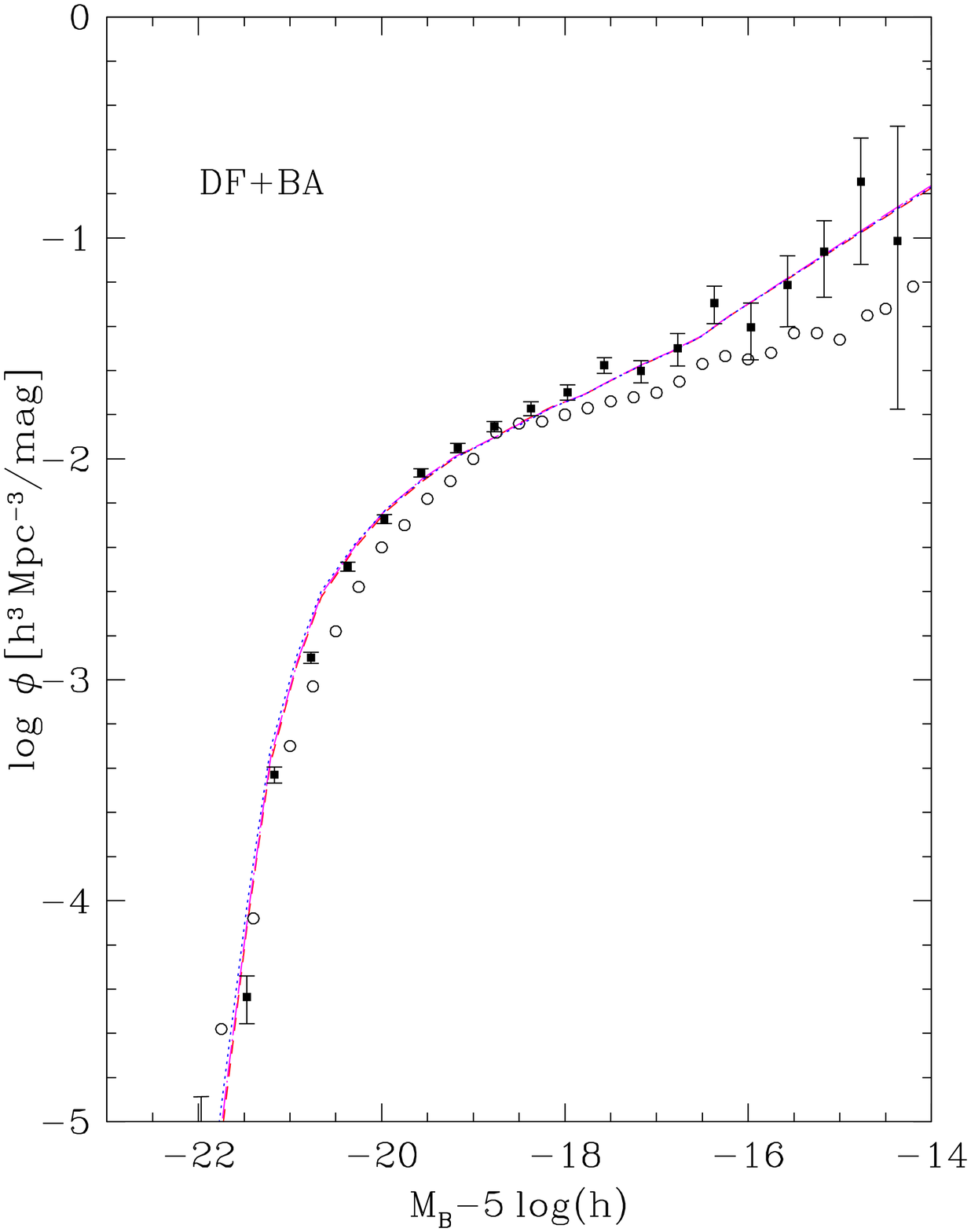}}} 
\end{center}
{\footnotesize
Fig. 6 -  
The local LF computed with the DF process alone (left panel) and with 
the inclusion of satellite aggregations (DF+BA, right panel). The different 
curves (almost overlapping here) refer to different choices of the dust 
extinction curve; MW (dashed line), SMC (dot-long dash), and Calzetti (dotted). 
The data are taken from Zucca et al. (1997, filled squares) and from the 2dFGRS 
survey (Madgwick et al. 2001, open circles).
\vspace{0.3cm}}

The low-$z$ LFs presented in fig. 6 show how satellite aggregations flatten the 
LF at faint/intermediate luminosities. Such an effect is purely dynamical, so 
that the agreement of the model with the Tully-Fisher relation shown in fig. 5 
is unchanged. At the very faint end, note the upturn of the predicted local LF 
in the DF+BA case, as expected from our discussion in \S 3.4.  The local trend 
toward a flatter LF shown by the DF+BA model persists at $z\sim 1$ (see fig. 7), 
where our model actually provides a better fit to the shape of the data 
distribution when compared to the DF case.

The high-$z$ LFs are compared with data in the bottom panels of fig. 7. Note 
that all models overpredict the number of faint galaxies as noticed in Poli et 
al. (2001). Several processes not yet properly inserted in the SAMs can be at 
the origin of the discrepancy. As discussed in sect. 4, simple variations of the 
stellar feedback prescritions adopted in the model are not a viable solution, 
being constrained by the observed Tully-Fisher relation. On the other hand, our 
treatement of binary aggregations considerably {\it reduces} the discrepancy, as 
shown by the right-hand panels in fig. 7. A residual excess of the predicted 
over the observed LF remains at the faint end. 

\vspace{0.1cm}
\begin{center} \scalebox{0.8}[0.65]{\rotatebox{0}{\includegraphics
{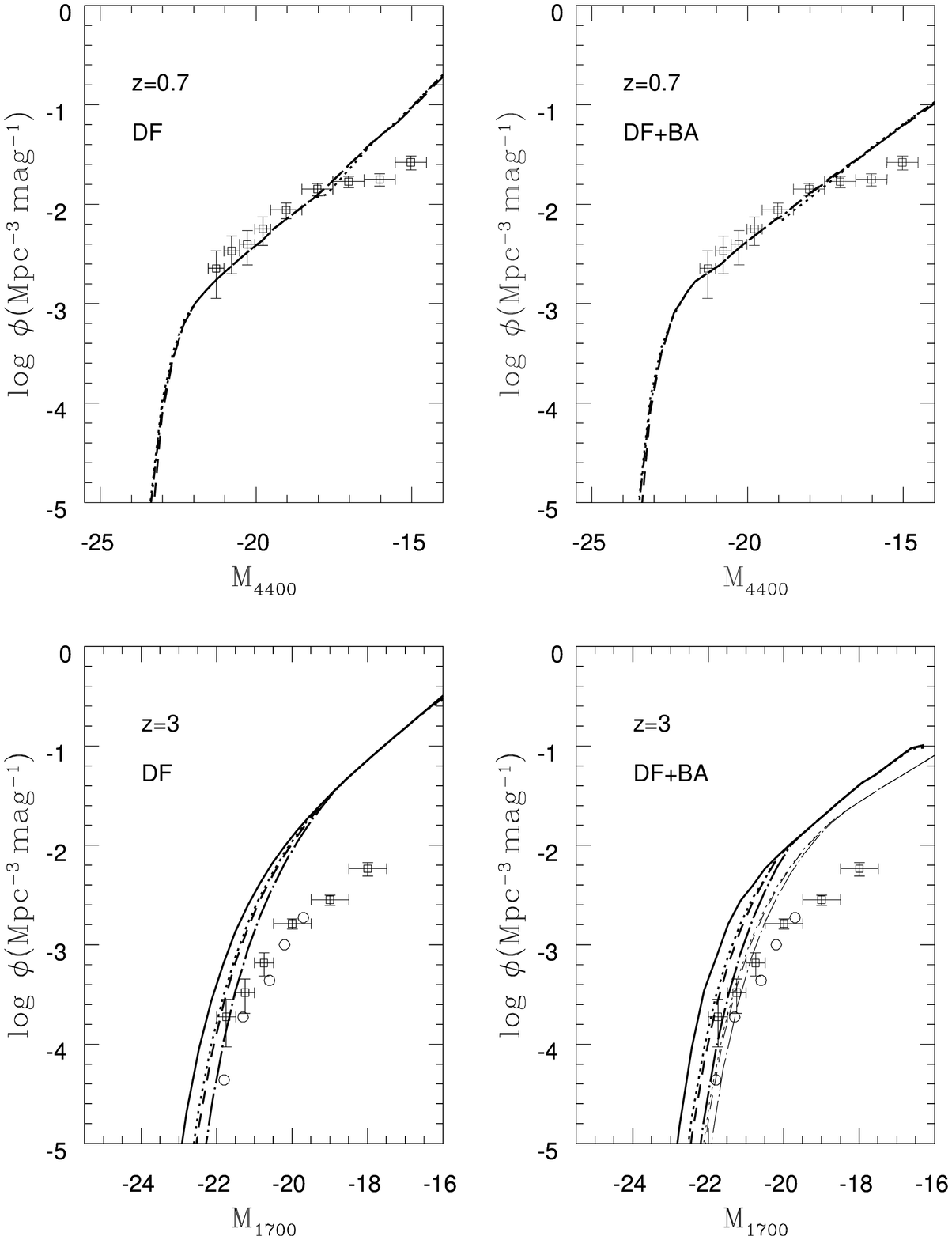}}}
\end{center}
{\footnotesize
Fig. 7 - 
The LF in the B-band at $z=0.7$ (top panels) and at 1700 ${\AA}$ at 
$z=3$ (bottom panels) for the DF process alone (left panels) and with the 
inclusion of satellite aggregations (DF+BA, right panels). The 
solid curves refer to the LF without dust extinction, while the other 
curves refer to different choices of the dust extinction curve: 
MW (dashed line), SMC (dot-long dash). and Calzetti (dotted). 
The data with the error bars correspond to the LF computed from photometric 
redshifts by Poli et al. (2001); the data corresponding to the spectroscopic 
survey by Steidel et al. (1999) are also shown by the circles in the bottom 
panels. In the right-bottom panel we also show the effect of adopting the Sheth 
\& Tormen (1999) instead of the Press \& Schechter expression for the halo mass 
distribution; the corresponding galaxy LFs (derived including DF+BA) are 
represented as thin curves (the line types correspond to the different extinction 
curves as above).
} \vspace{0.5cm}

A possible origin for it can be found in the statistics of the DM haloes (the 
Press \& Schechter mass distribution) hosting the galaxies. To investigate the 
issue we computed the LF from our model (including the binary aggregations of 
satellite galaxies) adopting the Sheth \& Tormen (1999) form (recalled in the 
Appendix A) for the mass function of the DM haloes hosting the galaxies; this is 
widely held to provide a better description of the halo statistics at high 
redshifts compared to the canonical Press \& Schechter distribution. 

The result (see right-bottom panel in fig. 7) is to further reduce the excess to 
a factor $\sim 2.5-3$ for luminosities fainter than $M_{1700}\approx -19$. At 
present, two explanations may be offered to account for the residual excess.

First, incompleteness in the data. This has been estimated by Vanzella et al. 
(2001) in the HDF-S data; at the faintest limits used to compute the LF 
($I_{AB}=27.25$), it has been found to be close to 1.6 for  extended sources. 
Even applying this correction factor to 
the whole faintest bin of the observed LF, the excess still remains.

Second, additional sources of feedback; e.g., the feedback due to the 
photoionization of the intergalactic medium by photons escaping with high 
efficiency from stars and 
quasars (Benson et al. 2001) could reconcile the predictions with the observed 
LFs. In this context, the contribution of binary aggregations reduces the amount 
of feedback required to yield flat LFs as to match the observations.

\section{DISCUSSION, CONCLUSIONS, AND PERSPECTIVES}

We have developed a semi-analytic model (SAM) for galaxy evolution, which 
includes the effects of binary aggregations between satellite galaxies 
orbiting inside common DM haloes (see fig. 1). The corresponding evolution of 
the galaxy mass function (see the velocity functions in fig. 3) has been 
calculated in a $\Lambda$-CDM cosmology by solving in each host halo a kinetic 
Smoluchowski equation  grafted onto our SAM code. 

The main effects of aggregations on the galaxy {\it mass distribution} are 
summarized as follows:

i) Starting at $z\approx 3$ (see fig. 4), aggregations of satellite galaxies 
gradually {\it deplete} the number of low/intermediate mass galaxies, and {\it 
flatten} the slope of the mass function (see fig. 3).  The aggregation cross 
section strongly depends on the ratio $v/V_{rel}$ between the galaxy circular 
velocity and the velocity dispersion of the halo wherein it orbits, so the 
increase of the average $V_{rel}$ following from the hierarchical clustering 
results in a progressive shift of the range of circular velocities where the 
aggregations flatten the mass distribution. At $z=0$ the resulting circular 
velocity distribution is flatter for small/intermediate circular velocities ($50 
\lesssim v\lesssim 200$ km/s, see fig. 3). The result agrees with the outcomes 
from N-body simulations.

ii) The total (dynamical friction plus binary aggregations) net merging rate 
{\it peaks} at $z\approx 3.5$ and {\it declines} sharply afterwards, in 
agreement with N-body results and consistent with observations. The contribution 
of satellite aggregations to the destruction rate (which mainly affects the low 
mass end of the mass distribution) becomes comparable to that produced by 
dynamical friction for $z< 1$, thus sustaining the galaxy destruction rate at 
low redhsifts. We have checked that the construction and destruction rates 
corresponding to the dynamics in our model (see fig. 4) agree with those 
obtained from N-body simulations within their limited mass resolution. 
 
The flattening of the mass distribution at small/intermediate masses 
velocities $50\lesssim v \lesssim 150$ km/s) affects the galaxy luminosity 
functions as shown in figs. 6, 7. These, obtained on adopting for the stellar 
section  the standard prescriptions of SAMs, have been compared with data from 
deep surveys with spectroscopic or photometric redshifts. The results concerning 
the {\it luminosity functions} are summarized as follows:

$\bullet$ at low redshifts, the binary aggregations {\it flatten} the LF at 
small/intermediate luminosities, producing also an {\it upturn} at very faint 
magnitudes $M_B>-16$. At magnitudes brighter than $L_*$ the LFs are not 
appreciably affected.

$\bullet$ at higher redshifts, the excess of the standard SAM predictions 
over the observed LFs at the faint end (by a factor $\sim 8$ at 
$z\approx 3$ for $M_{1700}\approx -18$, see Poli et al. 2001) is 
considerably {\it reduced} by the introduction of satellite aggregations 
computed in our present model. At $z\sim 3$, our LFs computed  
adopting the Sheth \& Tormen (1999) mass distribution for the host 
DM haloes further reduces the excess by a factor $\approx 3$ in the 
faintest bins ($-19.5\lesssim M_{1700}\lesssim -18$).

The origin of the residual excess at high redshifts requires further 
investigation. Incompleteness at the faintest limits of the observed LF 
($I_{AB}=27.25$) has been found to be around 1.6 for point and extended sources 
(Vanzella et al 2001). With regard to curing the residual excess, 
we note that changing 
the parameters in the canonical stellar feedback prescriptions is not a viable 
solution, since a larger feedback parameter (as required to flatten 
the galaxy LF still more) would result in a Tully-Fisher relation that is 
too faint when compared to the data. An alternative may be provided by sources of 
feedback other than SNe, e.g., the feedback due to photoionization of the 
intergalactic medium by stars and quasars (as advocated by Benson et al. 2001) 
could reconcile the predicted with the observed LFs. In this context, the 
contribution of binary aggregations to the  flattening of the LF will alleviate 
the need for extreme feedback efficiencies. Further insight could be provided 
by the comparison of the model predictions with K-band luminosity functions and 
with the baryonic mass function of galaxies (as suggested by Salucci \& Persic 
1999), a point we will address elsewhere. 

On the other hand, our conclusions about the faint end of the galaxy LFs are 
little affected by other processes recently implemented in the SAMs, namely, 
the additional starbursts associated with the mergers following satellite 
aggregations (Somerville \& Primack 1999), and the effect of tidal stripping of 
galactic subhaloes inside the host DM haloes (Benson et al. 2001). The former 
authors associate starbursts (with tunable timescale and amplitude) with each 
newly formed merger resulting from binary aggregations. We did not implement 
them in our model in order to single out dynamical effects of our description 
without introducing new (and uncertain) free parameters in the model. On the 
other hand, the starbursts associated with the mergers resulting from the galaxy 
encounters are expected to have a only a moderate inpact on the LF at 
faint/intermediate luminosities.  In fact, the dynamics described by eqs. (3) 
and (4) is such that the aggregation events mainly occur between intermediate 
and small mass galaxies (the first being favoured by their larger cross 
section and the latter by their large number), to form larger units; thus, the 
{\it destruction} term dominates at the small/intermediate masses while the {\it 
construction} term dominates at large masses. The starbursts associated with 
mergers are then expected to brighten mainly the bright end of the LF.

As for tidal stripping,  we note that Benson et al. (2001) have shown that it 
plays a minor role in determining the shape of the global galaxy luminosity 
function. This process, if anything, affects mainly the mass distribution at the 
very low-mass end (circular velocities $v\approx 20-50$ km/s), while satellite 
aggregations modelled here affect the mass distribution at larger scales 
($v\approx 50-150$ km/s). 

We stress that our description of the dynamics of galaxies inside common DM 
haloes does {\it not} introduce new free parameters in the SAMs; so, it 
constitutes a step toward improving the physical treatement of the processes 
involved in galaxy formation and evolution,  rather than merely enlarging the 
set of phenomenogical scaling laws and the associated list of free paremeters. A 
further step in such a direction will be constituted by a refined description of 
how central dominant galaxies form. In fact, although in the present paper the 
formation is treated in terms of dynamical friction to retain continuity with 
other works, in principle it could be included in the statistical description 
provided by the Smoluchowski kinetic equation. The process is due to the rare 
slow encounters for which the focussing term $v/V_{rel}$ contained in eq. (4) 
dominates.  In this case, the $V_{rel}$-dependent cross section (i.e., eq. 4 
without the velocity average) would become more than linear in mass in the low 
velocity tail of $g(V_{rel})$ (in particular, $\Sigma\propto m^{4/3}$ holds in 
such a case). In these conditions the aggregation dynamics described by eq. (3) 
are marked by a {\it  phase transition} of gravitational nature which breaks the 
system of colliding galaxies, originally comprised of a number of comparable 
objects, into two phases: a prompt, dominant  merger; and a number of satellites 
of much smaller mass, which gradually aggregate to the dominant object (see 
Cavaliere, Colafrancesco \& Menci 1991; Menci, Colafrancesco \& Biferale 1993). 
The transition takes place at time around $\tau_{dyn}\,(m/M)^{5/6}$ (the 
detailed expression is given by the above authors), which turns out to be 
comparable with the timescale actually adopted in the present paper, see eq. (2).  

Thus, in principle, the entire dynamics of galaxies in common haloes, including 
satellite aggregations and formation of central dominant mergers, could be 
treated within SAMs using only the kinetic description given in \S 3.2. In 
practice, however, this would require the detailed consideration of the galaxy 
velocity distribution to describe properly the probability for the occurence of 
slow, resonant encounters leading to the phase transition; this goes beyond the 
{\it average} description provided by eq. (4). In the present paper the 
formation of a central merger is described only in terms of dynamical friction; 
this can be viewed as a {\it mean field}, average rendition of the critical 
phenomenon leading to the phase transition. In perspective, we plan to describe 
all galaxy interactions in terms of the kinetic theory; this will be the subject 
of a future paper. 
\acknowledgments{
We gratefully acknowledge constructive advice of our referee toward 
correcting our manuscript and improving our presentation. Work supported by 
partial grants from ASI and MIUR. 
}

\newpage
\section*{APPENDIX A}
\setcounter{equation}{0} \renewcommand{\theequation}{A-\arabic{equation}}

The statistics of the virialized DM condensations in the framework of the 
hierarchical clustering is commonly described in terms of the Extended Press \& 
Schechter Theory (EPST; Bower 1991; Bond et al. 1991; Lacey \& Cole 1993). Here 
we summarize the results used in the text.

The number density of virialized structures of mass $M$ at the cosmic time $t$ 
is given by the  Press \& Shechter (1974) formula \begin{equation} 
N_H(M,t)=\sqrt{2\over\pi}\, {\rho_0\over M^2}\, \bigg|{d ln\sigma\over d ln 
M}\bigg|\, {\delta_c(t)\over\sigma(M)}\, e^{-
{\delta_c(t)^2\over2\,\sigma^2(M)}}~, 
\end{equation}
where $\rho_0$ is the cosmic average matter density, $\sigma (M)$ is the rms 
density fluctuations of the linear perturbation field in spheres of mass $M$ 
(see, e.g., Lacey \& Cole 1993), and $\delta_c(t)$ is the density threshold for 
collapse. For $\sigma(M)$ we adopt the form corresponding to Cold Dark Matter 
(CDM) density perturbations in a flat Universe dominated by a cosmological 
constant $\Omega_{\Lambda}=0.7$. The threshold $\delta_c(t)$ corresponds to the 
value -- extrapolated to the present using the linear growth factor 
$D(t,\Omega_0,\Omega_{\Lambda})$ for the density perturbations -- of the 
overdensity of a homogeneous sphere at the point where the exact non-linear 
theory predicts collapse to a singularity. Its normalization $\delta_{c0}$ at 
$z=0$ and its time evolution $D^{-1}(t)$ depend on cosmology; in a critical 
$\Omega=1$ Universe $\delta_{c0}=1.686$ and $D^{-1}(z)=(1+z)$ hold; the 
corresponding expressions for a flat Universe with a nonzero cosmological 
constant are given by, e.g., Eke, Cole \& Frenk (1996).

Since the variance of the density field is an inverse function of $M$ and the 
density threshold $\delta_c(t)$ lowers with time, larger and larger 
overdensities collapse to form structures of increasing mass (hierarchical 
clustering). For a given DM mass $M$ it is possible to compute the merging rates 
and the progenitor distributions.

In particular, the probability that a condensation of mass $M$ present at time 
$t$ form a halo of mass between $M' $ and $M' +dM'$ at time $t+dt$ is given by 
the following expression: 
\begin{equation} {d^2 P_H(M\rightarrow M',t)\over d 
M'\,dt} =  {1\over\sqrt{2\pi}}\, \Bigg[{\sigma^2\over\sigma'^2(\sigma^2-
\sigma'^2)}\Bigg]^{3/2}\,
e^{-{\delta_c^2(t)\,(\sigma^2-\sigma'^2)\over 2\,\sigma^2\,\sigma'^2}} \,
\Big|{d\sigma'^2\over dM'}\Big|\,\Big|{d\delta_c(t)\over dt}\Big|~, 
\end{equation}
where $\sigma$ and $\sigma'$ are the rms density fluctuations corresponding to 
the masse $M$ and $M'$, respectively. 

The history of the previous mergers experienced by a mass $M$ is instead 
described by the probability distribution that a given mass $M$ at time $t$ has 
a progenitor of mass $M'$ at time $t'<t$: 
\begin{equation} 
{d P_H\over dM'}(M',t'\rightarrow M,t)={\delta_{c}(t')-\delta_{c}(t)\over
(2\pi)^{1/2}(\sigma'^2-\sigma^2)^{3/2} }\,{M\over M'}\, 
\Big|{d\sigma'^2\over dM'}\Big|\; exp{
\Bigg\{-{[\delta_{c}(t')-\delta_{c}(t)]^2\over 2\,(\sigma'^2-\sigma^2) } \Bigg\} }~. 
\end{equation}

The survival time of a halo of mass $M$ is defined as the cosmic time at which 
its mass has grown to $q\,M$, and hence depends on the choice of the ``mass 
step'' of the merging tree. For a generic step parameter $q$, the probability 
$prob\big[\tau_l<\tau \big]$ that a halo with mass $M$ at time $t$ has a 
survival time $t+\tau_l$ with $\tau_l<\tau$ reads (Lacey \& Cole 1993): 
\begin{mathletters} \begin{eqnarray} prob\big[\tau_l<\tau \big]& = & {1\over 
2}\,{[\delta_c(t)-2\,\delta_c(t+\tau)]
\over \delta_c(t)}\,e^{2\,\delta_c(t+\tau)\,[\delta_c(t)-\delta_c(t+\tau)]
\over \sigma^2(M)}\,\nonumber\\
& \times & [1-erf(X)]+{1\over 2}\,[1-erf(Y)]\\
X & \equiv & {\sigma^2(q\,M)\,[\delta_c(t)-2\,\delta_c(t+\tau)]+
\sigma^2(M)\,\delta_c(t+\tau)\over
\{2\,\sigma^2(M)\,\sigma^2(q M)\,
[\sigma^2(M)-\sigma^2(q M)]\}^{1/2}}\\
Y & \equiv & {\sigma^2(M)\,\delta_c(t+\tau)-\sigma^2(q M)\,\delta_c(t)
\over \{2\,\sigma^2(M)\,\sigma^2(q M)\,
[\sigma^2(M)-\sigma^2(q M)]\}^{1/2}
}
\end{eqnarray}
\end{mathletters}
To make contact with previous works (Cole et al. 1994; Cole et al. 2000), our 
default choice for the mass step is $q=2$. We have verified that the SAM results 
are robust to smaller values down to $q=1.2$, as discussed by Cole et al. 
(2000).

Finally, we recall the expression for the mass function recently proposed by 
Sheth \& Tormen (1999; see also Sheth, Mo \& Tormen 2001) to provide a better 
fit to the N-body results. This has the form 
\begin{equation}
N_H(M,t)=A\,\sqrt{2a\over\pi}\,
{\rho_0\over M^2}\,
\bigg|{d ln\sigma\over d ln M}\bigg|\, 
\Bigg[1+\Bigg({\sigma^2\over a\,\delta_c^2(t)}\Bigg)^p\,\Bigg]\,
{\delta_c(t)\over\sigma(M)}\,
\,e^{-{a\delta_c^2(t)\over 2\,\sigma^2(M)}} ~.
\end{equation}
The values of the fitting coefficients $A=0.3222$, $a=0.707$ and $p=0.3$ are 
obtained by the above authors from comparison with the N-body results. The 
agreement of the above expression for $N_H(M)$ with the simulations has been 
confirmed by an independent analysis (Jenkins et al. 2001).

\section*{APPENDIX B}
\setcounter{equation}{0} \renewcommand{\theequation}{B-\arabic{equation}}

We derive the average value for the tidal radius $r_{tid}$ of galaxies 
orbiting inside a host halo with radius $R$ and circular velocity $V$. 
Such a radius is that appropriate for a galactic subhalo which survives 
the tidal stripping of the host halo, a condition which requires  
the density of the galactic subhalo within $r_{tid}$ to exceed 
the density of the host halo interior to the pericentre $r_p$ of its 
orbit. 

To this end we adopt the approach of Ghigna et al. (1998) who showed 
how the condition for survival against tidal stripping translates 
approximately into $r_{tid}\approx r_p\,(v/V)$. The above expression for 
$r_{tid}$ has been tested by the above authors against high resolution 
N-body simulations, and has been proved to agree with the values of 
$r_{tid}$ measured in the simulations for all subhaloes except the 
few on very eccentric orbits (the latter have measured radii 
larger than expected partly due to the formation of tidal tails). 

We average the above relation for $r_{tid}$ over the distribution of 
pericentres obtained by Ghigna et al. (1998) from N-body simulations, 
which we fit with a modified lognormal expression (in the variable 
$(r_p/R-0.08)$ with logarithmic mean -1.3 and logarithmic variance 0.6). 
Performing the average yields for $r_{tid}$ the following expression: 
\begin{equation}
\langle r_{tid}\rangle=R\,(v/V)\int_{r_{cut}}^{1}\,
\,p(r_p/R)\,d(r_p/R)\,(r_p/R)\equiv \alpha\,R\,v/V,  
\end{equation}
where the lower limit $r_{cut}$ corresponds to the minimum tidal radius 
that a galactic subhalo can have without being severely distorted or 
disrupted. Following Bullock, Kravstov and Weinberg (2000), we adopt for 
$r_{cut}$ the radius of the peak of the galaxy velocity profile. For a 
Navarro et al. (1997) circular velocity profile $r_{cut}=2.16\,r_{200}/c$ 
holds, where $c$ is the concentration parameter of the subhalo, and 
$r_{200}$ is the radius where the average density of the subhalo would 
equal $200\,\rho_c$. The equations relating $c$ and $r_{200}$ with the 
circular velocity $v$ of the subhalo are given by the above authors. The 
validity of the above value for $r_{cut}$ has been tested against N-body 
results by Bullock et al. (2000). 

It is easy to recast the relation $r_{tid}=\alpha\,R\,v/V$ in terms of 
density. Substituting the relation $v/V=(m/M)^{1/2}\,(R/r_{tid})^{1/2}$ 
one obtains $(r_{tid}/R)^3=\alpha^2\,m/M$ so that the ratio of the 
subhalo to the host halo average densities is given by $1/\alpha^2$. 
The average over the distribution of pericentres defining the value 
of $\alpha$ yields typical values for such a ratio ranging between 
3 and 5, depending on the concentration parameter $c$ of the subhaloes. 

Note that taking the tidal radius $r_{tid}=r_p\,v/V$ as the limiting 
radius of the subhaloes is fully appropriate only for subhaloes  
much smaller than their host halo. This is the case for the bulk of the 
population of satellite galaxies (the one mainly contributing to binary 
aggregations) for which our treatment constitutes a valid approximation.


\newpage

\begin{references}
\reference{}Adelberger, K.L., Steidel, C.C., Giavalisco, M., 
Dickinson, M., Pettini, M., Kellogg, M., 1998, ApJ, 505, 18 
\reference{}Benson, A.J., Lacey, C.G., Baugh, C.M., Cole, S., \& Frenk, C.S., 
2001, preprint [astro-ph/0108217] 
\reference{}Bond, J.R., Cole, S., Efstathiou, G., \& Kaiser, N., 1991, ApJ, 379, 440 
\reference{}Bower, R.J., 1991, MNRAS, 248, 332 
\reference{}Bruzual, A.G., \& Charlot, S., 1993, ApJ, 105, 538 
\reference{}Bullock, J.S., Kravtsov, A., V., Weinberg, D.H., 2000, 
ApJ, 539, 517
\reference{}Calzetti, D., 1997, in Proc. of "The Ultraviolet Universe 
at Low and High Redshift : Probing the Progress of Galaxy Evolution", 
AIP Conference Proceedings, 408, p.403
\reference{}Carlberg, R.G. et al., 2000, ApJ, 532, L1 
\reference{}Cavaliere, A., Colafrancesco, S., \& Menci, N., 1992, ApJ, 392, 41 
\reference{}Cavaliere, A., Colafrancesco, S., \& Menci, N., 1991, ApJ, 
376, L37
\reference{}Cavaliere, A., \& Menci, N., 1993, ApJ, 407, L9 
\reference{}Cavaliere, A., \& Menci, N., 1997, ApJ, 480, 132 
\reference{}Cole, S., Aragon-Salamanca, A., Frenk, C.S., Navarro, J.F., Zepf, 
S.E., 1994, MNRAS, 271, 781 
\reference{}Cole, S., \& Lacey, C.G., 1996, MNRAS, 281, 716 
\reference{}Cole, S., Lacey, C.G., Baugh, C.M., Frenk, C.S., 2000, MNRAS, 319, 168 
\reference{}Cross, N., et al., 2001, MNRAS, 324, 825 
\reference{}Eke, V.R., Cole, C.S., Frenk, C.S., 1996, MNRAS, 282, 263 
\reference{}Fontana, A., Menci, N., D'Odorico, S., Giallongo, E., Poli, F., 
Cristiani, S., Moorwood, A., \& Saracco, P., 1999, 310, L27 
\reference{}Ghigna, S., Moore, B., Governato, F., Lake, G., Quinn, T., \& 
Stadel, J., 1998, MNRAS, 300, 146 
\reference{}Giallongo, E., Menci, N., Poli, F., D'Odorico, S., Fontana, A. 2000, ApJ, 530, L73
\reference{}Giovanelli, R., Haynes, M. P., da Costa, L. N., Freudling, W., 
Salzer, J. J., \& Wegner G., 1997, AJ, 113, 22 
\reference{}Gnedin, O.Y., \& Ostriker, J.P. 1997, ApJ, 474, 223 
\reference{}Goodwin, S.P., Pearce, F.R., \& Thomas, P.A., 2000, preprint [astro-ph/0001180] 
\reference{}Kauffmann, G., White, S.D.M., \& Guiderdoni, B., 1993, MNRAS, 264, 201 
\reference{}Klypin, A., Kravtsov, A.V., Valenzuela, O., \& Prada, F., 
1999, ApJ, 522, 82
\reference{}Jenkins, A., Frenk, C.S., White, S.D.M., Colberg, J.M., Cole, S., 
Evrard, A.E., Couchman, H.M.P., \& Yoshida, N., 2001, MNRAS, 321, 372
\reference{}Lanzoni, B. 2000, PhD Thesis (IAP, Paris)
\reference{}Lacey, C., \& Cole, S., 1993, MNRAS, 262, 627 
\reference{}Le F\`evre, O., et al., 2000, MNRAS, 311, 565 
\reference{}Loveday, J. 1997, ApJ, 489, 29 
\reference{}Mac Low, M., \& Ferrara, A., 1999, ApJ, 513, 142 
\reference{}Madau, P., Pozzetti, L., \& Dickinson, M., 1998, ApJ, 498, 106 
\reference{}Madgwick, D.S., et al., 2001, submitted to MNRAS 
[astro-ph/0107197]
\reference{}Makino, J., \& Hut, P., 1997, ApJ, 481, 83 
\reference{}Mamon, G.A., 1992, ApJ, 301, L3
\reference{}Marzke, R.O., Huchra, J.P., \& Geller, M.J., 1994, 
ApJ, 428, 43
\reference{}Mathewson, D.S., Ford, V.L., \& Buchhorn, 1992, ApJS, 81, 413 
\reference{}Menci, N., Colafrancesco, S., \& Biferale, L. 1993, Journal de Physique, 3, 1105 
\reference{}Monaco, P. 2001, in Proc. of Vulcano Workshop {\it Chemical 
Enrichment of Intracluster and Intergalactic Medium}, F. Matteucci \& R. Fusco 
Femiano eds., in press 
\reference{}Mo, H.J, Mao S., \& White, S.D.M., 1998, MNRAS, 295, 319 
\reference{}Murali, C., Katz, N., Hernquist, L., Weinberg, D.H., Dav\'e, R., 
2001, preprint [astro-ph/0106282] 
\reference{}Navarro, J.F., Frenk, C.S., \& White, S.D.M. 1997, ApJ, 490, 493 
\reference{}Poli, F., Giallongo, E., Menci, N., D'Odorico, S., \& Fontana, A., 
1999, ApJ, 527, 662 
\reference{}Poli, F., Menci, N., Giallongo, E., Fontana, A., Cristiani, S., \& 
D'Odorico, S., 2001, ApJ, 551, L45; erratum, 554, L127
\reference{}Pozzetti, L., Madau, P., Zamorani, G., Ferguson, H.C., Bruzual A.G., 
1998, MNRAS, 298, 1133 
\reference{}Press, W.H., \& Schechter, P., 1974, \apj, 187, 425 
\reference{}Salucci, P., \& Persic, M., 1999, MNRAS, 309, 923
\reference{}Saslaw, W.C., 1985, {\it  Gravitational Physics of Stellar and 
Galactic Systems} (Cambridge: Cambridge Univ. Press) 
\reference{}Sheth, R.K., \& Tormen, G., 1999, MNRAS, 308, 119 
\reference{}Sheth, R.K., Mo, H.J., \& Tormen, G., 2001, MNRAS, 323, 1 
\reference{}Somerville, R.S., \& Primack, J.R., 1999, MNRAS, 310, 1087 
\reference{}Somerville, R.S., Primack, J.R., \& Faber, S.M., 2001, MNRAS, 320, 504 
\reference {}Steidel, C. C., Giavalisco, M., Pettini, M., Dickinson, M., \& 
Adelberger, K. L., 1996, ApJ, 462, L17 
\reference{}Steidel, C.C., Adelberger, K.L., Giavalisco, M., Dickinson, M., 
Pettini, M., 1999, ApJ, 519, 1 
\reference{}Steinmetz, M., \& Bartelmann, M., 1995, MNRAS, 272, 570 
\reference{}Sutherland, R., \& Dopita, M., 1993, ApJS, 88, 253 
\reference{}Taffoni, G., Mayer, L., Colpi, M., \& Governato, F., 2001, 
to appear in the Proceedings of the Conference "Chemical Enrichment 
of Intracluster and Intergalactic Medium", ASP conference Series 
[astro-ph/0109029]
\reference{}Taylor, J.E., \& Babul, A., 2000, preprint [astro-ph/0012305] 
\reference{}Tormen, G., 1997, MNRAS, 290, 411
\reference{}Sheth, R.K., \& Tormen, G., 1999, MNRAS, 308, 119 
\reference{}Sheth, R.K., Mo, H.J., \& Tormen, G., 2001, MNRAS, 323, 1 
\reference{}Smoluchowski, M., 1916, Phys. Z., 17, 557 
\reference{}Trubnikov, B.A., 1971, Soviet. Phys. Doklady, 16, 124 
\reference{}Vanzella E. et al., 2001, AJ, 122, 219 
\reference{}Warren, M.S., Quinn, P.J., Salmon, J.K, \& Zurek, W.H., 
1992, ApJ, 399, 405
\reference{}White, S.D.M.,  \& Frenk, C.S. 1991, ApJ, 379, 52 
\reference{}Willick, J.A., Courteau, S., Faber, S.M., Burstein, D., Dekel, A., 
\& Kolatt, T., 1996, ApJ, 457, 460 
\reference{}Wu, K.K.S., Fabian, A.C., Nulsen, P.E.J., 2000, MNRAS, 318, 889 
\reference{}Zucca, E., et al., 1997, A\&A, 326, 477 
\end{references}
\end{document}